\documentclass[aps,prr,twocolumn,superscriptaddress,floatfix,letterpaper,nofootinbib]{revtex4-1}

\usepackage[normalem]{ulem}

\usepackage{amsthm}
\theoremstyle{theorem}

\newtheorem{Proposition}{Proposition}

\newcommand{\w}{{\rm w}}
\newcommand{\SO}{{\rm SO}}
\newcommand{\Spin}{{\rm Spin}}
\newcommand{\U}{{\rm U}}
\newcommand{\SU}{{\rm SU}}

\newcommand{\bea}{\begin{eqnarray}}
\newcommand{\eea}{\end{eqnarray}}

\usepackage{graphicx}
\usepackage{amsmath}
\usepackage{amsthm}
\usepackage{amstext}
\usepackage{amssymb}
\usepackage{mathrsfs}
\usepackage{amsfonts}
\usepackage{amsbsy} 
\usepackage{dcolumn}
\usepackage{bm}
\usepackage{multirow}

\usepackage{color}
\usepackage[colorlinks,citecolor=blue,linkcolor=blue,urlcolor=blue]{hyperref}
\definecolor{red}{rgb}{1,0,0}
\definecolor{blue}{rgb}{0,0,1}
\definecolor{dblue}{rgb}{0,0,0.4}
\definecolor{green}{rgb}{0,1,0}
\definecolor{black}{rgb}{0,0,0}
\definecolor{white}{rgb}{1,1,1}

\definecolor{brn}{rgb}{.8,.4,.0}
\definecolor{redo}{rgb}{1,.5,.0}
\definecolor{ddgrn}{rgb}{0,0.4,0}
\definecolor{dgrn}{rgb}{0,0.55,0}
\definecolor{dbl}{rgb}{0,0,0.5}

\usepackage[bbgreekl]{mathbbol}
\usepackage{amscd}

\newcommand{\Z}{\mathbb{Z}}

\renewcommand{\v}[1]{\boldsymbol{#1}} 
 
\newcommand{\ii}{\hspace{1pt}\mathrm{i}\hspace{1pt}}
\newcommand{\ee}{\hspace{1pt}\mathrm{e}}

\newcommand{\<}{\langle} 
\renewcommand{\>}{\rangle} 

\newcommand{\Ref}[1]{Ref.~\onlinecite{#1}}
\newcommand{\Eq}[1]{(\ref{#1})} 
\newcommand{\eq}[1]{(\ref{#1})} 
\newcommand{\eqn}[1]{Eqn.\,(\ref{#1})}

\newcommand{\Tr}{{\rm Tr}}

\newcommand{\prt}{\partial}

\newcommand{\ie}{{\it i.e.~}}

\newcommand{\bpm}{\begin{pmatrix}}
\newcommand{\epm}{\end{pmatrix}}
\newcommand{\bmm}{\begin{matrix}}
\newcommand{\emm}{\end{matrix}}

\newcommand{\cV}{ {\cal V} }

\newcommand{\al}{\alpha} 
\newcommand{\bt}{\beta} 
\newcommand{\del}{\delta}

\newcommand{\ga}{\gamma} 
\newcommand{\Ga}{\Gamma} 
 
\newcommand{\la}{\lambda}

\newcommand{\si}{\sigma}

\newcommand{\nn}{{\nonumber}}

\def\Ext{\operatorname{Ext}}
\def\B{\mathrm{B}}
\def \A{\mathcal{A}}
\def \H{\operatorname{H}}
\def\Sq{\mathrm{Sq}}
\def\TP{\mathrm{TP}}

\usepackage{sseq}
\usepackage[all,cmtip]{xy}
\usepackage{tikz-cd}
\usepackage{tikz}
\usetikzlibrary{matrix}
\usetikzlibrary{decorations.markings}
\usetikzlibrary{positioning}
\usetikzlibrary{shadings}

\usepackage{datetime}

\usepackage{enumitem} 
\usepackage{moreenum}

\usepackage{forest}

\def\Ext{\operatorname{Ext}}
\def \Hom{\operatorname{Hom}}

\newcommand{\Sec}[1]{Sec.~\ref{#1}}

\begin{document}

\begin{titlepage}

\title{
A Non-Perturbative Definition of the Standard Models
}

\author{Juven Wang}
\affiliation{School of Natural Sciences, Institute for Advanced Study, Princeton, NJ 08540, USA }
\affiliation{Center of Mathematical Sciences and Applications, Harvard University, MA 02138, USA}

\author{Xiao-Gang Wen}
\affiliation{Department of Physics, Massachusetts Institute of
Technology, Cambridge, MA 
02139, USA}


\begin{abstract} 

The Standard Models contain chiral fermions coupled to gauge theories.  It has
been a long-standing problem to give such gauged chiral fermion theories a
 quantum non-perturbative definition.   By classification of quantum anomalies and
symmetric invertible topological orders via a mathematical cobordism theorem
for differentiable and triangulable manifolds, and the \emph{existence} of
symmetric gapped boundary for {the trivial symmetric invertible topological orders},
we propose that Spin(10) chiral fermion theories with Weyl fermions in 16-dimensional spinor representations can be defined on a 3+1D lattice,
and subsequently dynamically gauged to be a Spin(10) chiral gauge theory.  
As a result, the Standard Models from the 16n-chiral fermion SO(10) Grand Unification can
be defined non-perturbatively via a 3+1D local lattice model of bosons or
qubits. 
Furthermore, we propose that
Standard Models from the 15n-chiral fermion SU(5) Grand Unification can also be
realized by a 3+1D local lattice model of fermions.  


\end{abstract}


\maketitle

\end{titlepage}

{\small \setcounter{tocdepth}{1} \tableofcontents }

\section{Introduction and Definitions}

The \emph{Standard Models} \cite{G6179,W6764,SW6468}, 
gauge theories with the Lie algebra $u(1) \times su(2) \times su(3)$ in 3+1D,
coupled to fermions and bosons, are believed to describe elementary
particles.\footnote{
Elementary particles include fermions from quarks and
leptons, and bosons from gauge mediators and Higgs particle.  Gravitons are not
yet discovered experimentally. In addition, in our work, we do not consider any
dynamical gravity; we only consider anomalies of gauge or gravitational
non-dynamical background fields.  The local Lie algebra of standard {Standard
Models} is $u(1) \times su(2) \times su(3)$, but the global structure Lie group
can be $ \frac{\U(1) \times \SU(2)\times\SU(3)}{\Z_q}$ where $q=1,2,3,6$, see a
recent overview \cite{Tong2017oea1705.01853} on this issue. 
{In fact, as we will show later that for SO(10) and SU(5) Grand Unifications,
it is more natural to study the case $q=6$.}
Also, we denote the $d$-dimensional space and 1-dimensional time 
as $d+1$D.} 
In the standard
\emph{Standard Model}, there are 15 of 2-component complex Weyl fermions per family.
The $\SU(5)$ Grand Unification \cite{GG7438} has 15 complex Weyl fermions per family.
There are also non-standard Standard Models, such as the one from the $\SO(10)$
Grand Unification \cite{FM7593} which has 16 complex Weyl fermions per
family. But for a long time, the Standard Models were only defined via a perturbative
expansion, which is known not to converge.  So the Standard Models were not yet
known to be well-defined quantum theories.  This is related to the
long-standing gauged chiral fermion problem: How to define a chiral fermion
theory, with the parity violation \cite{PhysRev.104.254}, coupled to the gauge field,
{non-perturbatively and in the same dimension}, as a well-defined quantum theory with a
finite-dimensional Hilbert space for a finite-size system (for details, see
Appendix \ref{defs}), {but without suffered from fermion doublings \cite{Nielsen1981}}. In this work, we use the term \emph{gauged chiral fermion
theory} to mean chiral fermion theory coupled to a non-dynamical background
gauge field.  In fact, the gauge theories focused in this article are mostly
non-dynamical, unless mentioned otherwise.

{There were many previous pioneer attempts, such as a lattice gauge
approach \cite{K7959}, 
Ginsparg-Wilson fermion
approach \cite{GW8249}, Domain-wall fermion approach \cite{K9242,S9390}, and
Overlap-fermion approach \cite{NN9362,L9995}. In the Ginsparg-Wilson fermion
approach, the to-be-gauged symmetry is not strictly an onsite symmetry but only a quasi-local symmetry 
(see \ref{onsite} and \cite{CLW1141,CGL1314, Wang:2013yta, Wang:2017loc}, the quasi-local symmetry is still a non-onsite symmetry), thus it is much challenging to
gauge. (The abelian chiral gauge theory is achieved by Ref.~\cite{L9995}, however, the non-abelian case is still an open question.) 
In the Domain-wall fermion approach, we have an extra dimension,
where the dynamical gauge fields can propagate. 
The Overlap-fermion approach is a reformulation of the Domain-wall
fermion approach. The above approaches normally start with a spacetime Euclidean lattice path integral
and implement the Ginsparg-Wilson fermion.} 

{In contrast, in our work, we do \emph{not} formulate a spacetime lattice path integral \emph{nor} Ginsparg-Wilson fermion. 
Instead, we consider a discretized spatial lattice Hamiltonian with a continuous time, with additional criteria (see \ref{localLatt}): 
(1) with a tensor product Hilbert space, (2) with all interaction terms bounded by a finite range of lattice spacings (called short-range interactions), 
(3) we only discuss onsite symmetries (see \ref{onsite}).
Below we refer to our setup as a \emph{local lattice model}.\footnote{
{For a concrete lattice model, we mostly focus on a spatial lattice Hamiltonian. However, our arguments and Propositions \ref{p1}, \ref{p2} and \ref{p3}
are more general than a Hamiltonian picture, they are also applicable to quantum field theory and spacetime path integral approaches.}
}}

In this work, we aim to show 
{nontrivial evidence} that the gauged chiral fermion problem in both the
16n-fermion and the 15n-fermion Standard Models can be solved via a generalized
lattice gauge approach under \emph{local lattice model} (\ref{localLatt}).  In the standard lattice gauge approach, the fermions
do not interact directly.  The generalized lattice gauge approach simply adds
an extra direct fermion interaction or an indirect fermion interaction via
some Higgs fields.  A generalized lattice gauge approach, called 
 the mirror fermion approach, was proposed in 1986 \cite{EP8679,M9259}. In such
an approach, 
one starts with a 
lattice model containing chiral fermions (the normal
sector) and a chiral conjugated mirror sector (the mirror sector), with a to-be-gauged symmetry
acting as an onsite symmetry.  Then, \emph{one includes a proper fermion
interaction} \cite{S8456,S8631} \emph{in such a local lattice model, attempting to gap
out the mirror sector completely, without breaking the onsite symmetry and
without affecting the low energy properties of the normal sector}.  {This
is the key step, which will be referred to as \emph{gapping out the mirror sector
without breaking the (to-be-gauged) symmetry}.} Last, one can gauge the onsite symmetry to
obtain a gauged chiral fermion theory, regularized by a 
{local lattice model}.  

\Ref{EP8679} proposed a way to gap out the mirror sector without breaking the
symmetry, by introducing composite fermion fields formed by mirror fermion
fields, and 
{by adding symmetric mass terms between composite fermion
fields and the mirror fermion fields to make all those fermion fields formally
massive.  However, such a proposal cannot work in general. Even we can make all
the fermion fields formally massive, it does not imply we can fully gap out the
mirror sector.  This is because, even for some models with a \emph{perturbative local
anomaly},\footnote{
{We overview the concepts of anomalies, including
\emph{perturbative local
anomaly} and 
\emph{non-perturbative global anomaly} in Appendix \ref{ano}. In the main text, however, 
we purposefully reduce the information on anomalies but focus on the
mathematically much well-defined concepts called the \emph{cobordism theory}.}
\label{footnote:anomalies}} 
one can find composite fermion fields formed by mirror fermion fields
and to make all those fermion fields formally massive (see the Appendix of the
arXiv version of \Ref{W1301}).}  Also, the extensive studies on the previous mirror
fermion proposal \cite{GPR9396,L9418,GS9409013,CGP1247} 
{had difficulties} to demonstrate that interactions can fully gap out the mirror sector without breaking the
symmetry and without modifying the low energy dynamics of the normal sector.
It was pointed out in \Ref{BD9216} that ``attempts to decouple lattice fermion
doubles by the method of Swift and Smit cannot succeed.''  Consequently, many
people gave up the mirror fermion approach.

Recently, \Ref{W1301} 
{\emph{conjectured}} a new gapping condition for the
mirror sector:\\

\noindent
{\bf Proposition i}:
\emph{Consider a mirror sector in $d+1$D with a symmetry group $G$.  The mirror sector
can be gapped out without breaking the symmetry if (1) there exist (possibly
$G$-symmetry breaking) mass terms that make all the fermions massive; and (2)
$\pi_n(G/G_\text{grnd})=0$ for $n\leq d+2$, where $G_\text{grnd}$ is the
unbroken symmetry group.}\\

\noindent
The above \Ref{W1301}'s claim is obtained based on the following assumption (not
rigorously proven so far):\\ 

\noindent
{\bf Proposition ii}:
\emph{A $d+1$D $G$-symmetric non-linear $\si$-model with topologically trivial target
space $M=G/G_\text{grnd}$ (\ie $\pi_n(M)=0$ for $n\leq d+2$) allows a
gapped symmetric ground state.}\\

\noindent
Applying the above Propositions, \Ref{W1301} \emph{claimed} that 3+1D Spin(10) chiral
fermion theory with Weyl fermions in a 16-dimensional spinor representation 
can be defined via an \emph{interacting} local lattice model with a Spin(10) onsite
symmetry which can be gauged.\footnote{In this work, a \emph{local lattice
model} is a lattice model of bosons and/or fermions with short-range
interactions and a tensor-product structured Hilbert space, see (\ref{localLatt}).} The 16-fermion
Standard Model (\ie SO(10) Grand Unification or SO(10) Grand Unified Theory
$\equiv$ SO(10) GUT) can then be obtained from a 3+1D Spin$(10)$ chiral gauge
theory, coupled to Spin$(10)$ chiral Weyl fermions in the 16-dimensional
representation of Spin$(10)$.\\


\noindent \emph{Purpose of our present work}: {The homotopy group argument
in \Ref{W1301} only \emph{proposed} a sufficient condition.  There are mirror sectors
that do not satisfy the condition, that can still be gapped out without
breaking the symmetry 
{and without altering low energy physics in the normal sector}.  In this work, we are going to prescribe  a more general
condition, to capture the cases missed by \Ref{W1301}:

\begin{Proposition}
\label{p1}
Consider a mirror sector in $d+1$D with a symmetry group $G$.  Assume the mirror
sector can be viewed as a boundary of a 
{gapped local lattice model} in one
higher dimension with an onsite symmetry $G$.  If the 
{gapped local lattice model} in one higher dimension represents a trivial cobordism invariant (which corresponds to the identity element 0 in the cobordism group), 
then 
{the mirror sector can be chosen to be a fully gapped boundary.
}
\end{Proposition}
\noindent
Our above 
{statement} used the following assumption: 
\begin{Proposition}
\label{p2}
A symmetric gapped fermionic state 
{for a local lattice model}, with a trivial cobordism invariant,
must exist a fully gapped boundary that does not break the symmetry.
{By existence, it means that the  gapped boundary can be constructed.}
{If the boundary happens to be gapless,
the gapless modes can be 
fully gapped out by appropriate symmetry-preserving non-perturbative interactions without breaking the symmetry and without causing ground state degeneracy 
(neither symmetry breaking degeneracy, nor topological degeneracy \cite{KitaevKong2012.1104.5047, Wang2012am1212.4863, Kapustin2013nva1306.4254, Lan2014uaaLanWangWen1408.6514})}.
\end{Proposition}
\noindent
{To clarify, a symmetric gapped fermionic state of Proposition \ref{p2}
can also have another gapless boundary to host the normal sector.
The existence of a fully gapped boundary in Proposition \ref{p2} is meant for the mirror sector.}
\begin{itemize}[leftmargin=3mm]
\item 
{The existence of a symmetric gapped boundary above (in \ref{p1} and \ref{p2}) is believed to be true, because the
 trivial bulk with a trivial cobordism invariant is in fact the same phase of quantum matter as the trivial gapped vacuum.
Therefore the bulk gap, the boundary gap, and 
the trivial gapped vacuum in the exterior outside the bulk, 
all can be deformed to each other without closing the gap. See
Appendix \ref{ano} and \ref{sec:app-symmetric-gapped-boundary} for further details.}
\end{itemize}
Using the above {statements}, we will show that a 3+1D Spin(10) chiral fermion
theory with Weyl fermions in a 16-dimensional spinor representation can be
defined via an \emph{interacting} local lattice model with a Spin(10) onsite symmetry
which can be gauged.  In addition, we will show that a 3+1D SU(5) chiral fermion
theory with Weyl fermions in 5-dimensional and 10-dimensional representations
can be defined via an \emph{interacting} local lattice model with an SU(5) onsite
symmetry which can be gauged.  }


Last, we remark that to fully characterize the global symmetry in a
fermion system, we need to specify the full internal global symmetry group
$G_f$ and how the fermion number parity $\Z_2^f$ is embedded in $G_f$.  So we
can denote the fermion symmetry as $G_f \supset \Z_2^f$.  In our case, the full
internal symmetry is actually $G_f=\Spin(10)$, while SO(10) is the quotient
group $\Spin(10)/\Z_2^f =\SO(10)$.  So in this work, we use the name Spin(10)
chiral fermion model, rather than SO(10) chiral fermion model (which was
sometimes used by others).

\section{Cobordism Theory and Symmetric Gapped Boundary}
\label{sec:Cobordism}

Let us first explain the cobordism theory used in Proposition~\ref{p1} and \ref{p2}.
Based on a theorem of Freed-Hopkin \cite{FH160406527} 
and an extended generalization \cite{W2, HAHSII1912.13504,HAHSIII1912.13514}
(including higher symmetries \cite{GW14125148,K032,W0303,LW0316,Y10074601,B11072707,NOc0605316,NOc0702377}) 
there is a 1-to-1 correspondence between  
``the deformation class of 
invertible topological quantum field theories (iTQFTs\footnote{
{It is called an invertible TQFT because its partition function ${\bf Z}(M^D)$ on any closed manifold $M^D$ must have its absolute value $|{\bf Z}(M^D)|=1$,
namely ${\bf Z}(M^D)=\ee^{\ii \theta}$ can only be a complex phase. }
{On a closed spatial manifold $M^{D-1}$, it always has a single ground state ${\bf Z}(M^{D-1} \times S^1)=1$ with no topological ground state degeneracy.
Thus, 
${\bf Z}(M)=\ee^{\ii \theta}$ has an inverted phase ${\bf Z}^\dagger(M)=\ee^{-\ii \theta}$ that can be defined as its complex conjugated iTQFT.
The combined iTQFT ${\bf Z}(M) \cdot {\bf Z}^\dagger(M) =1$ is the trivial iTQFT (\ie the trivial gapped vacuum).}}) 
\cite{KW1458,F1478} with symmetry (including higher symmetries)'' 
and ``a cobordism group.''\footnote{By all symmetric iTQFTs, 
their classifications and characterizations depend on the category of manifolds that can detect them. The
categories of manifolds can be: TOP (topological manifolds), PL (piecewise linear manifolds), or DIFF (differentiable thus equivalently smooth manifolds), etc. 
These categories are different, and they are related by the inclusions:
\bea
\rm{TOP} \supseteq \rm{PL}  \supseteq \rm{DIFF}.
\eea
In contrast, triangulable manifolds are smooth manifolds at least for dimensions up to $D=4$ (\ie the ``if and only if'' statement is true below $D\leq 4$). 
The concept of piecewise linear (PL) and smooth DIFF structures are equivalent in dimensions $D\leq 6$.
Thus all symmetric iTQFT classified by the cobordant properties of smooth manifolds have a triangulation (thus a lattice regularization) on a simplicial complex (thus a UV competition on a lattice). 
This implies a correspondence between ``the symmetric iTQFTs (on smooth manifolds)'' and ``the symmetric invertible topological orders (on triangulable manifolds)''
for $D\leq 4$.
This leads to our application of this mathematical fact on the lattice regularization 
of symmetric iTQFTs and symmetric  invertible topological orders for various Standard Models of particle physics.
In this work, we only focus on the smooth differentiable (DIFF) manifolds and
their associated all possible {iTQFTs}.
The tools we
use in either case would be a certain version of cobordism theory suitable for
a specific category of manifolds. 
}
More precisely, there is a 1-to-1 correspondence (isomorphism ``$\cong$'') between the following two well-defined ``mathematical objects'' (these ``objects'' turn out to form 
the abelian group structures):
\bea \label{eq:thm}
&&\left\{\begin{array}{ccc}\text{Deformation classes of the reflection positive }\\ 
\text{$D$-dimensional extended invertible}\\
\text{topological field theories (iTQFT) with} \\
\text{symmetry group }
G={\frac{{G_{\text{spacetime} }} \times  {{G}_f }}{{N_{\text{shared}}}}}
\end{array}\right\}\nn \\
&&\cong[MT(G),\Sigma^{D+1}I\Z]_{\text{tors}}.
\eea
The $MT(G)$ is the Madsen-Tillmann spectrum \cite{MadsenTillmann4} of the $G$ group,
the $\Sigma$ is the suspension, the $I\Z$ is the Anderson dual spectrum, and 
the $\Sigma^{D+1}I\Z$ is the ${D+1}$-th suspension of the spectrum.
The ${\text{tors}}$ means taking only the finite group sector (\ie the torsion group).
The right-hand side is the torsion subgroup of the homotopy classes of maps from a Thom-Madsen-Tillmann spectrum \cite{thom1954quelques, MadsenTillmann4} to
a shift of the Anderson dual to the sphere spectrum.
{The spacetime symmetry ${G_{\text{spacetime} }}$ and the internal symmetry ${{G}_f }$, mod out the shared common normal subgroup
${N_{\text{shared}}}$, is combined to a $G$ structure: 
\bea  \label{eq:G}
G={\frac{{G_{\text{spacetime} }} \times  {{G}_f }}{{N_{\text{shared}}}}}.
\eea
This also means the pertinent iTQFTs of \Eq{eq:thm} are defined on manifolds with $G$ structure.}

In condensed matter physics, this \emph{roughly} means that 
\begin{Proposition}
\label{p3}
There is  a 1-to-1 correspondence \cite{FH160406527} between ``the invertible
topological order with symmetry (including higher symmetries 
{\cite{W2}})'' that can be regularized on a lattice in
its own dimensions and ``group {elements} as generators in a cobordism group,'' at least at
lower dimensions.
\end{Proposition}
{There is a logic gap here to establish Proposition~\ref{p3}, 
since by \Eq{eq:thm}, we only know there is a  1-to-1
correspondence between ``the iTQFTs with symmetry'' and ``the cobordism invariants from a cobordism group.'' 
We do not yet know if there is a  1-to-1
correspondence between ``the lattice invertible topological order with
symmetry'' and ``the iTQFTs with symmetry.''  In particular, we do not mathematically and rigorously prove how
to construct a lattice Hamiltonian realization for each iTQFT with symmetry classified by
a cobordism group.}  (We remark that some of the ``lattice invertible
topological orders with symmetry on a lattice'' are also called the Symmetry
Protected Topological/Trivial states (SPTs) \cite{GW0931,CLW1141,CGL1314, Senthil2014ooa1405.4015}, if they can deform to a
trivial tensor product state under \emph{local unitary transformations} after explicitly breaking the symmetry.)
{Regardless of a logic gap in the rigorous mathematical sense, 
the broad literature suggests strong physical evidence that
\begin{enumerate} [leftmargin=3mm, label=\textcolor{blue}{(\alph*)}:, ref={(\alph*)}]
\item 
\label{(a)} 
The classification of iTQFT  \cite{Kapustin2014tfa1403.1467, KTT1429, FH160406527, GW171111587, W2}
so far matches with
the classification of lattice invertible topological orders and lattice SPTs \cite{Senthil2014ooa1405.4015, Wang2014lca1401.1142, WG170310937, Wang2018pdc1811.00536}.
Many such iTQFTs can thus be constructed on the lattice Hamiltonian.
\item 
\label{(b)} 
Moreover, in \Eq{eq:thm}, we only focus on iTQFTs definable on 
differentiable and triangulable manifolds, thus those iTQFTs may be regularized by the same lattice 
from the simplicial complex of triangulable manifolds.
\end{enumerate}
In summary, based on the support of \ref{(a)} and \ref{(b)}, 
below we propose and assume that a refined and rigorous version of
Proposition~\ref{p3} is true.
}

{Proposition~\ref{p2} can be obtained from Proposition~\ref{p3}.  There can be two kinds
of gapped fermion systems on a lattice, those with topological excitations (which
may be fractionalized) and those without topological excitations (\ie all the
excitations correspond to the original fermions or bosons).  
%
{The gapped states with topological excitations, by definition, are the lattice \emph{non-invertible} topological orders.}
The gapped states without topological excitations are the ``lattice \emph{invertible}
topological orders with symmetry.''  According to Proposition~\ref{p3}, if a
``lattice invertible topological order with symmetry'' has a trivial cobordism
invariant, then it must be a ``trivial lattice invertible topological order
with symmetry.''\footnote{A ``trivial lattice invertible topological order
with symmetry'' means the ``trivial gapped vacuum'' in quantum field theory, 
or namely the ``trivial gapped featureless insulator'' in condensed matter.} 
 In other words, there exists a symmetry preserving local
unitary transformation that deforms the ``trivial lattice invertible topological
order with symmetry'' into a ``trivial tensor product state with
symmetry'' \cite{CGW1038}, 
{where its gapped symmetric boundary can always be constructed. 
(We provide more steps along with these logical arguments in Appendix \ref{props}.)}
Crucially, this is precisely why the cobordism approach allows us to obtain the gapping condition for the
mirror sector.}

{
{Proposition~\ref{p1} can be obtained from Proposition~\ref{p2}, if we can show that
the normal sector or the mirror sector can be realized as some boundary states
of a 
{gapped local lattice model}. We 
will provide such a local lattice model construction for the Spin(10) chiral fermion theory in \Sec{so10} and in \Sec{bLat} as an example.
}
}

{Although we propose Proposition \ref{p1}, \ref{p2}, and \ref{p3},
we do not require the complete versions of all these Propositions to establish our claim of a local lattice model with \emph{a chiral fermion low-energy spectrum}.
We only require the weaker Proposition \ref{p1}, let us clarify:
\begin{itemize}[leftmargin=3mm]
\item Proposition~\ref{p1} 's ``the existence of a fully gapped boundary'' is a \emph{static} statement.
Proposition~\ref{p2}'s ``the gapless sector can be fully gapped out without breaking the symmetry'' 
is a \emph{dynamic} statement, more changeling than Proposition~\ref{p1}.
But the two statements are related, their detailed relations are given in Appendix \ref{props} and in \ref{sec:app-symmetric-gapped-boundary}
based on physical intuitions of phase boundaries and quantum phase transitions.
In fact, we only require the weaker \emph{static} statement in Proposition~\ref{p1}'s ``the existence of a fully gapped boundary'' in order to establish the gapped mirror sector.
\item For Proposition \ref{p3},
we only require a local lattice construction for the cobordism class whose boundary gives rise to the normal sector or mirror sector.
To establish Spin(10), Spin(18), and SU(5) chiral fermion theories, we only require a local lattice construction of 
the trivial cobordism class (the identity element 0 in the cobordism group).
They happen to be a trivial bulk gapped insulator that we certainly can construct their local lattice model with a gapless normal sector on the boundary (\Sec{so10}).
\end{itemize}
}

In the following sections, we also provide the physics interpretations of the classifications of all 
{4+1D} {iTQFTs}
whose boundaries are associated with the
{3+1D} Spin(10) and Spin(18) chiral fermion theories (for SO(10) and SO(18) GUTs) in \Sec{so10},\footnote{{To be precise, \label{footnote:SM1} in order to embed the
standard Standard Model-like spacetime-and-internal symmetry group with this 
$G={\frac{{G_{\text{spacetime} }} \times  {{G}_f }}{{N_{\text{shared}}}}}$ structure \Eq{eq:G}
to the SO(10)
Grand Unification's $\frac{\Spin(D) \times \Spin(10)}{{\Z_2^f}}$, 
it is natural to
consider an alternative standard Standard Model spacetime-and-internal group $\Spin(D) \times
\frac{\U(1) \times \SU(2)\times\SU(3)}{\Z_6}$, 
while their gauge Lie algebra is still $u(1) \times su(2) \times su(3)$.  Here Spin(10) and $\frac{\U(1)
\times \SU(2)\times\SU(3)}{\Z_6}$ are their gauge groups respectively.}
See more discussions in footnote \ref{footnote:SM321}.
}  
and the {3+1D} SU(5) chiral fermion theories (for SU(5) GUTs) in \Sec{SU5}.
We relegate the mathematical calculation details on algebraic topology in Appendix \ref{Appendix:Cobordism}. (See also
\Ref{WW1910.14668}.)


\section{Spin$(N)$ Chiral Fermion Theory, and SO(10) and SO(18) Grand Unification}
\label{so10}

We now construct a local lattice Hamiltonian model.
{A 3+1D two-component  Weyl fermion described by a Hamiltonian (\ref{m1} defined in Appendix \ref{defs})
\begin{align}
\label{WF}
 H = \psi^\dag \ii \si^i \prt_i \psi, \ \ \
\si^{1,2,3} \text{ are Pauli matrices},
\end{align}
can be realized on the boundary of a fermion
hopping model on a 4D spatial cubic lattice with a Hamiltonian
operator \cite{W1301}
\begin{align}
 \hat H_\text{hop}=\sum_{\v i\v j} (t_{\v i\v j}^{ab} \hat c^\dag_{a,\v i} \hat c_{b,\v j} +h.c.),
\end{align}
which has 4 fermion orbitals ($a,b=1,\cdots ,4$) per site ($\v i, \v j$ for sites).  
The $h.c.$ contains the hermitian conjugate term.
The $4\times 4$
hopping matrices $t_{\v i \v j}$ are given by 
\begin{align}
\label{H4d}
&\ \ \ \
 H_{\text{4D}}( k_1, k_2, k_3, k_4)
\\
&=
2[\Ga^1\sin(k_1) +\Ga^2\sin(k_2) +\Ga^3\sin(k_3) +\Ga^4\sin(k_4)]
\nonumber\\
&\ \ \ \
+2\Ga^5[\cos(k_1) +\cos(k_2) +\cos(k_3) +\cos(k_4)-3]
\nonumber
\end{align}
in the momentum $\v k$-space, where 
$\Ga^1=\si^1\otimes \si^3$,
$\Ga^2=\si^2\otimes \si^3$,
$\Ga^3=\si^3\otimes \si^1$,
$\Ga^4=\si^0\otimes \si^2$, 
and $\Ga^5=\si^0\otimes \si^3 $,
which obey $ \{\Ga^i,\Ga^j\}=2\del_{ij} $.  If the 4D lattice is formed by
two layers of 3D cubic lattices, the one-body Hamiltonian in the
$(k_1,k_2,k_3)$-space is given by the following 8-by-8 matrix
\begin{align}
 H_{\text{3D}}&(k_1,k_2,k_3)
= \begin{pmatrix}
 M_1 &  M_2\\
 M_2^\dag &  M_1
\end{pmatrix},\ \ \ \ \text{where}
\nonumber\\
 M_1 &= 2[\Ga^1 \sin(k_1) +\Ga^2 \sin(k_2) +\Ga^3 \sin(k_3)]
\nonumber\\
& \ \ \ \ \ \
+2\Ga^5 [ \cos(k_1) +\cos(k_2) +\cos(k_3) -3],
\nonumber\\
 M_2 &= -\ii \Ga^4 +  \Ga^5
.
\end{align}
One can directly check that the above 3D fermion hopping model gives rise to a
two-component massless complex Weyl fermion on each of the two {3D} surfaces of the 4D
lattice.  The  Weyl fermion on one boundary is a left-hand Weyl fermion and on
the other boundary is a right-hand Weyl fermion. We have a similar result when
the 4D lattice is formed by many layers of 3D cubic lattices.
}

{The 16 copies of the local lattice model \eq{H4d} give rise to the 3+1D Weyl
fermions in the 16-dimensional spinor representation of the $\Spin(10)$ on the
lattice boundary's low energy spectrum.  The ground state of the 4+1D hopping model is 
\begin{itemize}[leftmargin=3mm]
\item a ``lattice invertible topological state (invertible topological order whose low energy is an iTQFT)
with a $\Spin(10)\supset \Z_2^f$ symmetry,'' since
it has no non-trivial topological excitations.  
 \item
a lattice non-trivial 4+1D $\Spin(10)$ \emph{non-interacting free} fermionic SPT
state \cite{Schnyder2008tyaRyu0803.2786, K0986, Schnyder2009Ryu0905.2029}. 
\end{itemize}
But such a state may correspond to 
\begin{itemize}[leftmargin=3mm]
\item
a trivial state for
$\Spin(10)$ SPT state in lattice \emph{interacting} fermionic SPT
systems \cite{GW1441,KTT1429,GK150505856,KT170108264,WG170310937}
and to a trivial cobordism class (in \ref{p1}, \ref{p2} and \ref{p3}) \cite{FH160406527, W2}.
\end{itemize}
If so, the
4+1D hopping model can have a symmetric gapped boundary, and the 3+1D $\Spin(10)$
chiral Weyl fermions (\ref{WF}) for the mirror sector can be gapped by
interactions without breaking the symmetry by {Proposition}
\ref{p1} and \ref{p2}.}

To show that the 4+1D hopping model gives rise to a trivial $\Spin(10)$ SPT state
in \emph{interacting} fermion systems, we use a recent {conjectured}
complete classification of interacting fermionic invertible topological orders
\cite{KTT1429, FH160406527, KT170108264, GW171111587,WY180105416, 181211959, W2}
with onsite symmetry, via {a twisted version of the spin cobordism theory} of Freed-Hopkin
\cite{FH160406527}.  This classification includes all known interacting fermionic SPT states and all known interacting fermionic invertible topological orders
on a lattice \cite{GW1441, GK150505856, WG170310937,LW180901112}.

We first note that, for fermions with the full symmetry $G_f\supset \Z_2^f$ in
the $D$ spacetime dimensions, they transform as $G=\frac{\Spin(D) \times
G_f}{{\Z_2^f}}$ under the combined spacetime symmetry ${G_{\text{spacetime}
}}=\Spin(D)$ rotation and the internal $G_f$ transformation, where a double-counted
fermion parity symmetry ${\Z_2^f}$ is mod out.  This shared normal subgroup
${\Z_2^f}$ is due to the fact that rotating a fermion by $2 \pi$ in the
spacetime (namely, the spin-statistics) gives rise to the same fermion parity
minus sign for the fermion operator $\psi \to -\psi$.

{To classify the iTQFT whose boundary can have a 3+1D Spin(10) chiral fermion theory, 
we focus on the following cobordism group
\begin{equation}
\Omega^{D=5}_{\frac{(\Spin(D=5) \times \Spin(10))}{\Z_2^f}} 
\equiv
{\mathrm{TP}}_{D=5}({\frac{(\Spin(D=5) \times \Spin(10))}{\Z_2^f}}). \quad\quad
\end{equation}
More generally,} we find that 
4+1D fermionic invertible topological orders with $G_f=\Spin(N) \supset \Z_2^f$ onsite global symmetry
for $N\geq 7$ are classified by the $5$-th cobordism group \cite{W2}:
\bea \label{eq:SpinSpinN}
\Omega^{D=5}_{\frac{(\Spin(D=5) \times \Spin(N))}{\Z_2^f}}  = \Z_2, \quad \quad N\geq 7.
\eea
Beware that we \emph{define} the cobordism group, classifying symmetric fermionic invertible topological orders, as
\bea \label{eq:cob-TP}
\Omega^{D}_{G} &\equiv&
\Omega^{D}_{({\frac{{G_{\text{spacetime} }} \times  {{G}_{f} }}{{N_{\text{shared}}}}})} \nonumber\\
&\equiv&
{\mathrm{TP}}_D(G)\equiv[MT(G),\Sigma^{D+1}I\Z], 
\eea
which stands for the homotopy classes of maps from Thom-Madsen-Tillmann spectrum \cite{thom1954quelques, MadsenTillmann4} 
$MT(G)$ to 
the ${D+1}$-th suspension of the Anderson dual spectrum
$\Sigma^{D+1}I\Z$.
Our notations follow Refs.~\cite{FH160406527, W2, GW171111587} and \cite{W2-1812.11968}:
TP abbreviates ``Topological Phases'' classifying the symmetric invertible topological orders (or invertible topological quantum field theories),
${N_{\text{shared}}}$ is the shared normal subgroup of ${G_{\text{spacetime} }}$ and ${{G}_{f} }$.

The cobordism group of topological phases (TP) defined in \cite{FH160406527} as 
$\TP_D(G) \equiv  \Omega^{D}_{G}$ classifies the deformation classes of reflection positive invertible $d$-dimensional extended topological field theories with symmetry group $G_D$. The $\TP_D(G)$ and the bordism group $\Omega_D^G$ are related by a short exact sequence
\begin{equation} \label{eq:Ext-extension-main}
\hspace{-4mm}
0\to\Ext^1(\Omega_D^G,\Z)\to\TP_D(G)\equiv  \Omega^{D}_{G} \to\Hom(\Omega_{D+1}^G,\Z)\to0,\quad\quad
\end{equation}
with Ext denotes the extension functor, see Appendix \ref{Appendix:Cobordism}.

{In contrast}, we \emph{do not} define the cobordism group as 
the usual definition of Pontryagin dual of the torsion subgroup ($\equiv$ {tors}) of the bordism group $\Omega^{D}_{G}$ as the homomorphism (Hom) map to U(1): 
\bea \label{eq:hom}
\mathrm{Hom}(\Omega^{D,\mathrm{tors}}_{G}, \mathrm{U(1)}),
\eea
although the torsion (\ie finite group) sectors of \eq{eq:cob-TP} and \eq{eq:hom} are equivalent.
Mathematical details for the above result are presented in
Refs.~\onlinecite{W2, GW171111587, WW1910.14668}.\footnote{In contrast, \Ref{Garcia-Etxebarria:2018ajm} computes
a different bordism group
$\Omega_{D=5}^{{(\Spin(D=5) \times \Spin(10))}}$ $=0$. Instead, we study
the bordism group $\Omega_{D=5}^{{(\Spin(D=5) \times \Spin(10))}/{\Z_2^f}}=\Z_2$, whose manifold generator can 
detect the new SU(2) anomaly \cite{WWW}.}  We classify
the deformation classes of invertible topological quantum field theories 
(further precisely, the reflection positive invertible extended topological field theories)
via $\Omega^{D}_{G}$, by classifying the cobordant differentiable and triangulable
manifolds with a stable $G$-structure, via associating them to the homotopy
groups of Thom-Madsen-Tillmann spectra \cite{thom1954quelques,MadsenTillmann4},
thanks to a theorem in \Ref{FH160406527}.


To be precise, here the spin cobordism theory {is 
believed} to
completely classify all the fermionic {iTQFTs}.
By applying this spin cobordism theory, 
we classify 4+1D $\Spin(N)$ SPT states and 4+1D $\Spin(N)$ symmetric invertible fermionic topological orders.
In fact, in this context, the 3+1D $\Spin(N)$ fermion theories already include all possible 3+1D $\Spin(N)$ chiral fermion theories that we need. 
To this end, we will especially focus on the 3+1D Spin(10) chiral fermion theories with Weyl fermions in a 16-dimensional spinor
representation.

The above $\Z_2$ classification in \eq{eq:SpinSpinN} implies that there is only one non-trivial 4+1D invertible fermionic
topological order with a $\Spin(N)$ onsite symmetry.  We find that such a
topological phase is characterized by a 5-dimensional topological invariant \cite{W2} 
written in terms of a bulk partition function
on a 5-manifold $M^5$,
\begin{align}
\label{topinv}
{\bf Z} = \ee^{\ii \pi \int_{M^5} \w_2(TM) \cup \w_3(TM)},
\end{align}
where $\w_n(TM)$ is the $n^\text{th}$-Stiefel-Whitney
class for the tangent bundle of $4+1$D spacetime manifold
$M^5$, and the $\cup$ is the cup product (which we may omit writing  $\cup$) \cite{milnor1974characteristic}.
We note that on $M^5$, we have a $\frac{\Spin(D=5) \times
\Spin(N)}{{\Z_2^f}}$ connection --- a mixed gravitational and gauge connection,
rather than a pure gravitational $\Spin(D=5)$ connection, {such that
$\w_2(TM)=\w_2(V_{\SO(N)})$ and $\w_3(TM)=\w_3(V_{\SO(N)})$, where
$\w_n(V_{\SO(N)})$ is the $n^\text{th}$-Stiefel-Whitney class for an ${\SO(N)}$
gauge bundle}.\footnote{In the context of anomalies (see Appendix
\ref{ano} for details), the boundary of this 4+1D $\Spin(N)$-SPT state may 
have a mixed anomaly of $\SO(N)$-gauge
bundle and spacetime geometry/gravity, {and we can use the this  3+1D anomaly on
the boundary to detect the bulk invertible topological order.}  
Namely, we find that \emph{there is only one possible candidate of 
the 3+1D anomaly for interacting fermion systems with a $\Spin(N)$ symmetry} ($N\geq 7$), 
which is a non-perturbative global mixed gauge-gravity (\ie gauge-diffeomorphism) anomaly
characterized by \Eq{topinv}.
}  
Thus, $M^5$ may not be a spin manifold (note that a spin manifold requires $\w_2(TM)=0$).  


{We can detect the 4+1D cobordism invariant $\ee^{\ii \pi \int_{M^5}
\w_2(TM)\w_3(TM)}$ for the 4+1D invertible fermionic topological order by study
its bondary state. In particular, if the 4+1D state has a boundary described by
3+1D $\Spin(N)$ chiral Weyl fermion theory, then we can detect the 4+1D
cobordism invariant via the $\Spin(N)$ representation of the chiral Weyl
fermions on the boundary.  Here we use a fact that the  4+1D cobordism
invariant can be detected by restricting to a $\SU(2)=\Spin(3)$ subgroup of
$\Spin(N)$ \cite{WWW}:  Let $n_j$ be the number of isospin-$j$
representations of $\SU(2)= \Spin(3) \subseteq \Spin(N)$ for
3+1D boundary chiral Weyl fermions, then
the 4+1D cobordism invariant $\ee^{\ii \pi \int_{M^5} \w_2(TM)\w_3(TM)}$ is
absent if
\begin{align}
\label{njnj}
\sum_{r=0}^\infty n_{2r+\frac12} \in \Z_{\text{even}},
\ \ \ \
\sum_{r=0}^\infty n_{4r+\frac32} \in \Z_{\text{even}}.
\end{align}
}

To see
how the representation of $\Spin(N)$ reduces to the representations of
$\SU(2)=\Spin(3)$, let us describe the representation of $\Spin(N)$ (the spinor
representation of $\Spin(N)$), assuming $N=$ even. We first introduce
$\ga$-matrices $\ga_a$, $a=1,\cdots,N$:
\begin{align}
 \ga_{2k-1}&=
\underbrace{\si^0\otimes \cdots \otimes \si^0}_{\frac{N}{2}-k  \ \si^0{\text{'s}}}
\otimes \si^1\otimes
\underbrace{\si^3\otimes \cdots \otimes \si^3}_{k-1 \ \si^3{\text{'s}}},
\nonumber\\
 \ga_{2k}&=
\underbrace{\si^0\otimes \cdots \otimes \si^0}_{\frac{N}{2}-k \ \si^0{\text{'s}}}
\otimes \si^2\otimes
\underbrace{\si^3\otimes \cdots \otimes \si^3}_{k-1  \ \si^3{\text{'s}}},
\end{align}
$k=1,\cdots,\frac{N}{2}$, which satisfy $ \{\ga_a,\ga_b\}=2\del_{ab}$ and
$\ga_a^\dag=\ga_a$.  Here $\si^0$ is the 2-by-2 identity matrix and $\si^l$ with
$l=1,2,3$ are the Pauli matrices.  The $\frac{N(N-1)}{2}$  hermitian matrices $
\ga_{ab}=\frac{\ii}{2}[\ga_a,\ga_b]= \ii \ga_a\ga_b, \  a<b$, generate a
$2^{N/2}$-dimensional representation of $\Spin(N)$.  The above
$2^{N/2}$-dimensional representation is reducible.  To obtain an irreducible
representation, we introduce
\begin{align}
\ga_\text{FIVE}&=(-\ii )^{N/2} \ga_1 \cdots \ga_{N}
=
\underbrace{\si^3\otimes \cdots \otimes \si^3}_{\frac{N}2 \ \si^3{\text{'s}}}.
\end{align}
We have $(\ga_\text{FIVE})^2=1$, its trace $\Tr (\ga_\text{FIVE})=0$, and
$\{\ga_\text{FIVE},\ga_a\}=[\ga_\text{FIVE},\ga_{ab}]=0$.  This allows us to
obtain two $2^{N/2-1}$-dimensional irreducible representations: one for
$\ga_\text{FIVE}=1$ and the other for $\ga_\text{FIVE}=-1$.

Now, let us consider an $\SU(2)=\Spin(3)$ subgroup of $\Spin(N)$, generated by
$\ga_{12} = I\otimes \si^0\otimes \si^3$, $\ga_{23} = I\otimes \si^1\otimes
\si^1$,  and $\ga_{31} = I\otimes \si^1\otimes \si^2$.  We see that the
$2^{N/2-1}$-dimensional representation of $\Spin(N)$ becomes $2^{N/2-2}$
isospin-1/2 representations of $\SU(2)$.   

{Summarizing the above results, we see that the $2^{N/2-1}$ copies of 4+1D
hopping model (\ref{H4d}) formed by many (but finite) layers of 3d cubic
lattices has a 3+1D boundary chiral Weyl fermion in the $2^{N/2-1}$-dimensional
representation of $\Spin(N)$ on one boundary (the normal sector), and a conjugate 3+1D chiral Weyl
fermion on the other boundary (the mirror sector).  For an even $N\geq 8$, the
3+1D boundary chiral Weyl fermions only reduces to an even number of
isospin-1/2 representations,  and, according to (\ref{njnj}), the 4+1D
cobordism invariant $\ee^{\ii \pi \int_{M^5} \w_2(TM)\w_3(TM)}$ is absent.
Thus the corresponding 4+1D bulk state is a trivial fermionic iTQFT (\ie the identity element
as the trivial cobordism class in {Proposition} \ref{p1} and \ref{p2}) with
a Spin(N) symmetry.  In this case, the mirror sector can be gapped out without
breaking the $\Spin(N)$ symmetry by introducing a proper symmetric fermion
interaction on the boundary ({Proposition} \ref{p1}).  Since the 4+1D hopping model has many layers and
the 3+1D boundary massless chiral Weyl fermion has no symmetric \emph{relevant} deformation 
operators (in the renormalization group sense, e.g., there is no $\Spin(N)$ symmetric mass term), the symmetric
interaction in the mirror sector on one boundary will not affect the low energy dynamics of the 3+1D
massless chiral Weyl fermion in the normal sector on the other boundary.}

%
Now we apply a well-known lattice method\\[1mm]
{{\bf Finite-width/layer lattice dimensional reduction:}
\bea
&&\text{\emph{An $n+1$D lattice model with finite layers along one}}\nn\\[-1mm]
&&\text{\emph{extra direction (a finite width $w$) can be dimensionally}}\nn\\[-1mm]
&&\text{\emph{reduced to an $n$D lattice model via absorbing the degrees}} \nn\\[-1mm]
&&\text{\emph{of freedom along w to the orbital in $n$D.}}  \label{eq:lattice-dimensional-reduction}
\eea}
Thus by \Eq{eq:lattice-dimensional-reduction}, 
the 4+1D hopping model with finite layers can be viewed as a 3+1D
lattice model with \emph{finite orbitals per site}, the  3+1D $\Spin(N)$ chiral Weyl fermion theory in the
$2^{N/2-1}$-dimensional representation can be regularized by a lattice model in
the same dimension without breaking the $\Spin(N)$ symmetry for even $N\geq 8$. 

In particular, for $N=10$, the Spin(10) chiral fermion theory with Weyl
fermions in a 16-dimensional spinor representation (similarly, for Spin(18)
chiral fermion theory in a 256-dimensional spinor representation) can be
{regularized} by a local lattice fermion model in the same dimension.  {After
regularizing the Spin(10) chiral fermion theory as a  lattice fermion model in
the same dimension (3+1D) with an onsite Spin(10) symmetry, we can gauge the onsite
Spin(10) symmetry
to obtain a gauged Spin(10) chiral fermion theory,\footnote{
{To gauge the onsite symmetry, 
one way is by inserting gauge variables on the 1-dimensional links between local sites.
This is known as the \emph{hard-gauge}, such that the outcome gauge theory does not have a tensor product Hilbert space (\ref{localLatt})
thus it is \emph{not} a local lattice model that we aim for.
However, we can further maintain a tensor product Hilbert space (\ref{localLatt})
by designing the \emph{soft-gauge}. We relegate the details of 
the \emph{soft-gauge} via a local lattice model in Appendix \ref{bLat}.
See discussions on \emph{hard-gauge} and \emph{soft-gauge} in \Ref{Wang:2017loc}.
}
 \label{footnote:hard-soft-gauge}
}
again regularized by a lattice model in the same dimension.  
}

%
%
%
%


We remark that,
in fact, for $N=3$, the 4+1D fermionic invertible topological orders with a
Spin(3)=SU(2) internal global symmetry are classified by the cobordism group of
${{(\Spin(D=5) \times \Spin(N=3))}/{\Z_2^f}}$ \cite{W2, WWW}: 
\begin{equation} \label{eq:SpinSpin3}
\Omega^{D=5}_{\frac{(\Spin(D=5) \times \Spin(3))}{\Z_2^f}} =\Omega^{D=5}_{\frac{(\Spin(D=5) \times \SU(2))}{\Z_2^f}}  = (\Z_2)^2.
\end{equation}
The corresponding cobordism invariant is given by
\begin{align}
\label{albt}
{\bf Z} = 
\ee^{\ii \al \pi \int_{M^5} \text{Arf} \, \; \, \w_3(TM)}
\ee^{\ii \bt \pi \int_{M^5}  \w_2(TM) \w_3(TM)}.  
\end{align}
Here Arf is the Arf invariant \cite{Arf1941},  {which characterizes the 1+1D
fermionic chain whose open ends host Majorana zero modes \cite{K0131}.} This
1+1D fermionic chain is also known as Kitaev chain \cite{K0131} whose low energy
physics is governed by a 1+1D invertible fermionic topological order.  The
above cobordism invariant can be detected by the SU(2) representations of 3+1D
boundary chiral Weyl fermions, and $\al,\bt$ in (\ref{albt}) are given
by \cite{WWW}
\begin{align}
\al=\sum_{r=0}^\infty n_{2r+\frac12} \text{ mod } 2,
\ \ \ \
\bt=
\sum_{r=0}^\infty n_{4r+\frac32}  \text{ mod }2.
\end{align}
In this work, we only suggest that there exists a
symmetric short-range non-perturbative interaction that can fully gap out the mirror sector
without breaking the $\Spin(10)$ symmetry.\footnote{
{There exists such a symmetric interaction preserving Spin(10). Moreover, for 16 chiral Weyl fermions at IR,
there is a U(16) global symmetry, with $\U(16) \supset \Spin(10)$. It is possible that additional constraints happen 
on what interactions we can engineer in the quotient space $\frac{\U(16)}{\Spin(10)}$ (also a homogeneous space)
without breaking {Spin(10)}.
We may need to preserve more than Spin(10) subgroup in the U(16).
}}  
Our approach only proves the
symmetric gapped boundary exists (via Proposition~\ref{p1} and \ref{p2}),  but {does
not} provide {a prescription to design} such an interaction. The approach in
\Ref{W1301,DW170604648} proposes a design: The interaction in the mirror
sector is given by the smooth orientation fluctuations of Higgs field (thus beyond
the Higgs mechanism \cite{BZ150504312,Wang:2017loc}), where a constant
orientation will gap out all the mirror fermions.  But the validity of the
design requires confirmation by numerical simulations.  A first step  is taken
in \Ref{DW170604648} for a 1+1D system. In such a design, crucially
the mass of the mirror fermions induced by the Higgs field must be comparable with
the fermion bandwidth.  
(Some other 
gapping-mirror-fermion approaches have also been {proposed \cite{Wang:2013yta,
YBX1451,YX14124784,BZ150504312, Wang:2018ugf}.})
Many previous calculations \cite{GS9409013,CGP1247}
checking the mirror fermion approach choose an induced energy gap (\ie an
effective mass) to be much bigger than the bandwidth (\ie at the infinite coupling
limit).
{The infinite coupling limit in the mirror sector generates a dead layer, 
a neighbor layer next to the mirror sector would become the new mirror sector with fermion doublings \cite{Nielsen1981},
which would fail to produce a chiral fermion/gauge theory at low energies.}

\section{SU(5) Chiral Fermion Theory and SU(5) Grand Unification}
\label{SU5}

Above we have discussed the lattice regularization of a $\Spin(10)$ gauged
chiral fermion theory.  To consider a lattice regularization of a $\SU(5)$
gauged chiral fermion theory (with $G_f=\Z_2^f \times \SU(5)$, but only SU(5)
will be gauged), we classify the 4+1D invertible fermionic topological order
with $G_f=\Z_2^f \times \SU(5)$ symmetry by a cobordism group defined in
\eqn{eq:cob-TP} (note that $\Spin(D=5)\supset \Z_2^f$) \cite{W2, WW1910.14668}:   
\begin{equation}
\label{eq:cobordism-SU5}
\Omega^{D=5}_{\Spin(D=5) \times \SU(5)}  
\equiv
{\mathrm{TP}}_D({\Spin(D=5) \times \SU(5)})
= \Z,
\end{equation}
where the topological invariant is given by the $\SU(5)$ Chern-Simons 5-form,
associated with \emph{perturbative local anomalies} captured by perturbative Feynman diagram
calculations in 3+1D.

%
%

{Again, such a cobordism invariant and the associated invertible
topological order can be detected by the boundary chiral fermions: if the 3+1D
boundary SU(5) chiral fermion theory is free from any of the $\Z$ class of SU(5)
perturbative local anomaly, then the corresponding  cobordism invariant and the 4+1D bulk invertible
topological order are trivial. } Thus, by {Proposition} \ref{p1} and \ref{p2}, \emph{any $\SU(5)$ gauged chiral fermion
theory that can be realized at the boundary of a 4+1D gapped local lattice
model, can be realized by a 3+1D local lattice model via the method \Eq{eq:lattice-dimensional-reduction}, provided that the
$\SU(5)$ gauge theory is free of the $\SU(5)$ perturbative anomalies} (see
also \ref{propoI} in Appendix \ref{defs}). In particular, the $\SU(5)$ grand unified
theory \cite{GG7438} can be regularized by a lattice.  This implies that its induced
15-fermion Standard Model can be regularized by a lattice fermion model.

\section{Implications and Conclusions}


{In fact, an $n+1$D G-symmetric iTQFT given by a cobordism class in Proposition~\ref{p1} and \ref{p2}
corresponds to an $n$D 't Hooft anomaly of $G$-symmetry
(see footnote \ref{footnote:anomalies} and the details of anomalies in Appendix \ref{ano}).
So a trivial cobordism class in $n+1$D for $G$-symmetry means
all-'t Hooft-anomaly-free in $n$D for the full $G$-symmetry.}
Namely, by far we only show that anomaly-free gauged chiral fermion theories can be
defined on a lattice with non-dynamical background gauge fields (\ref{m1} and
\ref{m2} in Appendix \ref{defs}), realized with onsite symmetries in its own
dimensions (via \Eq{eq:lattice-dimensional-reduction}).  However, we can obtain a dynamical chiral gauge theory (\ref{m3}
in Appendix \ref{defs}) by dynamically gauging the \emph{onsite} symmetry:
introducing dynamical gauge link variables between local sites (e.g.
dynamically sum over gauge inequivalent configurations in the partition
function) 
%
%
{ --- this is a \emph{hard-gauge} model but not a local lattice model, see footnote \ref{footnote:hard-soft-gauge}}.
{We can further apply the \emph{soft-gauge} method \cite{Wang:2017loc} to obtain a local lattice model, see Appendix \ref{bLat}.}
We  emphasize if all gauge invariant operators are bosonic, the
above dynamical lattice gauge theory coupled to fermions is actually a local
lattice bosonic model in disguise, as one can see from the
slave-particle/parton approach \cite{BA8880,AZH8845,DFM8826,LW0316,B0507}.

We remark that the dynamical $\Spin(10)$ chiral gauge theory coupled to Weyl
fermions in the 16-dimensional spinor representation is a local bosonic theory, since
all gauge invariant operators are bosonic.\footnote{In
Appendix \ref{bLat},
we provide the explicit slave-particle/parton
construction for a 4+1D \emph{local
bosonic} lattice model, whose boundary can give rise to the dynamical
$\Spin(10)$ chiral gauge theory coupled to Weyl fermions (\ref{m3}) in the 16-dimensional
representation.}   The lattice regularization that realizes the
dynamical $\Spin(10)$ chiral gauge theory is also a local bosonic model (see
Appendix \ref{bLat}).  In other words, the $\Spin(10)$ dynamical chiral gauge
theory with Weyl fermions in a 16-dimensional representation, and the induced
16-fermion Standard Model, can be realized as the low energy effective theory
of a local lattice model of qubits (since any local bosonic lattice model can
be viewed as a lattice model of qubits).  
{Based on the stability
of cobordism group of \eqn{eq:SpinSpinN} for $N \geq 7$, our result directly applies to
{a $\Spin(N=18)$
chiral gauge theory} \cite{Wilczek:1981iz, BZ150504312}, which is also a local bosonic
model.}
Thus our study implies that all elementary particles (except the graviton) can be viewed as originated from
qubits \cite{Wh01090120,W0303a,LW0622}. It is a concrete realization of ``it
from qubit \cite{W8954},'' representing an ultra unification of all gauge interactions
and matter fermions in term of quantum information (\ie qubits).  

The statement that all elementary particles arise from bosonic qubits has a
falsifiable experimental prediction: all fermions and their fermionic bound
states must carry non-trivial gauge charge \cite{LW0316,LW0510}. As a result,
the ``Standard Model'' from a lattice qubit model cannot just have a 
{$\frac{\U(1) \times \SU(2)\times \SU(3)}{\Z_q}$} 
gauge group, since such a Standard Model indeed has 
fermionic bound states that carry \emph{no} gauge charge.
Thus, the ``standard
model'' from a lattice qubit model must have a larger gauge group, e.g.\; adding a
new $\Z_2$ gauge sector,\footnote{See Footnote \ref{footnote:SM1},
\label{footnote:SM321}
we can show that
$$\hspace{-12mm}
\frac{\Spin(D) \times
\Spin(10)}{{{\Z_{q'}}}} 
\supset 
\Spin(D) \times \SU(5) 
\supset 
\Spin(D) \times \frac{\U(1) \times \SU(2)\times\SU(3)}{\Z_6},
$$
where $q'=1$ or $2$,
while 
$
\SO(10) 
\supset 
 \SU(5) 
\supset 
\frac{\U(1) \times \SU(2)\times\SU(3)}{\Z_6}
$ 
and 
$\Spin(10)\supset
 \SU(5) 
\supset 
\frac{\U(1) \times \SU(2)\times\SU(3)}{\Z_6}
$.
{When we gauge the [Spin(10)], 
we also require to gauge the $[\Z_2^f \times \SU(5)]$ in the embedded smaller group $\Spin(D) \times \SU(5)$.
The dynamically gauging $\Z_2^f$ symmetry produces the new $\Z_2$ gauge sector.}}   
where we gain a new cosmic string (whose spacetime trajectory is a 2-dimensional worldsheet) --- the
flux line of the new $\Z_2$ gauge field \cite{W1281}.

In contrast,  the dynamical $\SU(5)$ chiral gauge theory coupled to Weyl
fermions in the 5- and 10-dimensional representations is a fermionic theory definable on \emph{spin manifolds},
since some gauge invariant operators are fermionic.  The lattice model that
realizes the dynamical $\SU(5)$ chiral gauge theory is also a local fermionic
model (which is not a local lattice model of qubits).  The Standard Model from
local fermionic lattice models can have $\frac{\U(1) \times
\SU(2)\times\SU(3)}{\Z_q}$ as its gauge group, see Footnote
\ref{footnote:SM321}. It does not require extra gauge groups.

Lastly, we comment on the dynamics of these dynamical chiral gauge theories
(\ref{m3}, as highly long-range entangled states).  At the low energy of these
chiral gauge theories, there could be emergent symmetries (e.g. higher-form
symmetries \cite{Gaiotto:2014kfa} or higher symmetries in general \cite{W2}) having new 't
Hooft anomalies.  However, emergent new anomalies only mean the emergent
symmetries cannot be strictly regularized locally on-site, on-link,
on-$n$-simplex,  etc., which, we emphasize, is a rather distinct issue deviated
from regularizing chiral fermion theories which we solved earlier.  After
regularizing chiral fermion theories on a lattice, and after dynamically
gauging, the emergent new anomalies \emph{only} constrain the dynamics of gauge
theories (e.g. gapless near a quantum critical fixed point, or emergent
symmetry spontaneously broken, etc.).  We aim to address the dynamics of gauge
theories in future work.

\section{Acknowledgement}
We thank E.\,Witten for many helpful discussions and comments, and for a
collaboration on \Ref{WWW}.  JW thanks Z.\,Wan for related collaborations.  JW
is supported by Corning Glass Works Foundation Fellowship and NSF Grant
PHY-1606531.  XGW is partially supported by NSF grant DMS-1664412. This work is
also supported by NSF Grant PHY-1306313, PHY-0937443, DMS-1308244, DMS-0804454,
DMS-1159412 and Center for Mathematical Sciences and Applications at Harvard
University.

\appendix

\section{Definitions of Terminology and Discussions based on Anomalies}
\label{ano}

{
In the main text, we have described our results without directly mentioning the quantum anomaly (footnote \ref{footnote:anomalies}).
However, in literature, many people discuss the gauge chiral fermion problem in
terms of anomaly.  In this section, we will discuss our approach using the
concept of the anomaly.  We will carefully define several different anomalies.  
We will also carefully define several concepts of chiral fermion field
theory and the concepts of lattice theory as a well-defined quantum theory.
}

\subsection{Detailed definitions of some relevant concepts}
\label{defs}

We should clarify several related concepts of Spin(10) chiral fermion field
theories and models as follows:
\begin{enumerate} [label=\textcolor{blue}{Model \arabic*}:, ref={Model \arabic*}]

\item  \label{m1} \emph{Without gauging or before gauging} Spin(10) symmetry,
the theory is a ``{Spin(10) chiral fermion theory}'' 
with the full internal global symmetry $G_f=
\Spin(10) \supset {\Z_2^f}$.  In this case, we call the anomaly associated with
the global symmetry $G_f$ as the 't Hooft anomaly of $G_f$.  We classify
the \emph{'t Hooft anomaly} \cite{H8035} of $G_f$ in \eqn{eq:SpinSpinN} and \eqn{eq:cob-TP}.   

\item  \label{m2}
{We may twist the Spin(10) symmetry via a non-dynamical background
Spin(10) gauge field, known as the \emph{symmetry twist}. 
We name such a theory as a ``{Spin(10) gauged chiral fermion theory}.'' 
The {anomaly} of $G$ classified in \Eq{eq:SpinSpinN}
becomes the \emph{background gauge anomaly}, which is the same as the 't Hooft
anomaly in nature.}  

\item  \label{m3} 
\emph{After dynamical gauging} Spin(10) symmetry via dynamical weakly
fluctuating Spin(10) gauge field, the theory becomes a ``{Spin(10) chiral
gauge theory}.'' In this case, it is a standard terminology to call the
anomaly, descending from \ref{m1}'s 't Hooft anomaly and after gauging
Spin(10), as the \emph{dynamical gauge anomaly}.  We will thus also study the
dynamical gauge anomaly of $\Spin(10)$, thanks to \eqn{eq:SpinSpinN}. If any
model possesses any dynamical gauge anomaly, then this theory is inconsistent
thus ill-defined. 

\end{enumerate}

\Ref{W1301} {adopted} a new viewpoint (or a new definition) of anomalies proposed
in \Ref{W1313} (see \ref{b-anom}), for \emph{interacting quantum} theories,
which in turn leads to a classification of anomalies. Before we proceed, we
should clarify some conventions of terminology as the definitions:
\begin{enumerate}[label=\textcolor{blue}{Def.~\Roman*}:, ref={Def.~\Roman*}]

\item \label{localLatt}
\textbf{Well-defined quantum theories} are 
\emph{quantum theories} defined with 
{a finite-dimensional Hilbert space and}
{a finite-dimensional Hamiltonian matrix}
{for a finite-size system in the real space. 
In this work, we only focus on this class of {quantum theories.}}

{ \textbf{Local lattice models} are interacting or non-interacting
lattice models whose many-body Hilbert space $\cV$ has the following tensor
product decomposition
\begin{align}
\label{td}
 \cV=\bigotimes_i \cV_i
\end{align}
where $\cV_i$ is a finite-dimensional Hilbert space for each lattice site.}

By \emph{interacting} models, we mean the Hamiltonian contains certain higher-order terms beyond the quadratic terms
of fundamental lattice operators (such as quartic fermionic or spin operator terms beyond the quadratic terms).  

By \emph{non-interacting} models (or the so-called \emph{free} or \emph{quadratic} models),
we mean the Hamiltonian contains at most the quadratic terms (thus easily diagonalizable and solvable) 
of fundamental lattice operators.

{\textbf{Local interactions}: By local interactions, we mean
all interaction terms in the Hamiltonian (or a Lagrangian in the path integral) must be 
bounded by a \emph{finite range} of lattice spacings. We call these types as local, finite-range or
short-range interactions. We do \emph{not} allow infinite-range interactions,
\emph{nor} the interactions with strength exponentially-decay to zero only at infinite. 
For any interaction term of our lattice model, it must be bounded by a finite spatial range, say if the operators act on any site $i$ to $j$,
then the locality means that ``$|i-j| \leq $ a finite distance.'' 
}

We emphasize that conventional \emph{lattice gauge theories} with dynamical gauge fields are
usually \emph{not} local lattice models: 
Since there is a non-local gauge constraint,
thereby the tensor product decomposition
\eq{td} is violated. In this work, we do \emph{not} use 
models of conventional \emph{lattice gauge theories}, but limit ourselves to
only \emph{local lattice models}.

\item \label{onsite}
{For such local lattice models, the \textbf{onsite symmetry}
is defined as a global internal symmetry, whose symmetry transformation operator has the following  tensor product decomposition
\begin{align}
 U =\bigotimes_i U_i,
\end{align}
where $U_i$ is a unitary operator acting on $\cV_i$. }

\item \label{well-QFT}
\textbf{Well-defined quantum field theory} (living on the boundary of lattice model):  When we mention a ``well-defined''
quantum field theory (QFT), we always mean a \emph{limited class} of QFTs which
can be realized (\ie regularized) as the low energy effective boundary theory
of a gapped local lattice model (see \ref{localLatt}) in one higher dimension (so-called the bulk).
The global symmetry, if any, is realized as an onsite symmetry (see
\ref{onsite}) for the full bulk-boundary system. 
Such a QFT has at most the b-anomaly to be defined later in
\ref{b-anom}.  A ``well-defined''  QFT cannot have the r-anomaly to be defined
later in \ref{r-anom}. 

\item \label{anom-free}
\textbf{All lattice obstruction-free} (required to be regularizable in the same dimension):
 The above defined QFTs
include $d+1$D QFTs that can be realized by a lattice model in the \emph{same}
dimension $d+1$D (with the symmetry, if any, realized as an onsite symmetry or a local on-$n$-simplex symmetry\footnote{Here
we only focus on the well-defined $G$-symmetric QFTs 
with ordinary $G$-global symmetries (the 0-form symmetry in the sense of generalized global symmetries \cite{Gaiotto:2014kfa}).
If there is a generalized higher global symmetry \cite{Gaiotto:2014kfa}, then
we need to modify the ``lattice onsite symmetry realization'' to the ``lattice local on-$n$-simplex symmetry realization.''
\label{footnote:on-site-symmetry-on-simplex}
}), because the
gapped bulk in one higher dimension (of \ref{well-QFT}) can be a decoupled
gapped tensor product state.
This leads to a concept of 
\emph{all lattice obstruction-free}:
 By saying a $d+1$D QFT 
 is \emph{all lattice obstruction-free}, 
 we always mean a $d+1$D  well-defined quantum field theory in
\ref{well-QFT}, which can be realized as the low energy effective boundary
theory of a $d+2$D  gapped tensor product state (\ie a gapped trivial vacuum) on a
one-higher-dimensional lattice.  Note that a tensor product state (\ie a trivial state, with \emph{neither short-range nor long-range entanglement} of \ref{gapp-EE}) in a
local lattice model is defined as 
\begin{align}
 |\Psi\> = \bigotimes_i |\psi_i\>,\ \ \ \
|\psi_i\> \in \cV_i,
\end{align}
which can be gapped and decoupled from its boundary theory.
In contrast, the generic state is more general and is not necessarily a tensor product, such as
\bea
 |\Psi\> = \sum_{\{  \rm{c}_{\{ i\}} \} } \rm{c}_{\{ i\}} \Big( \bigotimes_i |\psi_i\> \Big),\ \ \ \
|\psi_i\> \in \cV_i,
\eea
with generic complex normalizable coefficients $\rm{c}_{\{ i\}}$.


\item \label{gapp-EE}
\textbf{Gapped system and entanglement:} 
{By a gapped tensor product state, we mean that the 
 tensor product state is a unique ground state (with an energy $E_0$) of some lattice bulk Hamiltonian system
 whose energy spectrum has 
 a finite energy gap $\Delta_E = E_{\text{excited}} - E_0 > 0$
 separated from all excited states $E_{\text{excited}}$.
 Below the energy gap $\Delta_E$, the system behaves as a gapped trivial vacuum (or a gapped trivial insulator in condensed matter) with no entanglement.}

{On the other hand, general gapped systems ($\Delta_E = E_{\text{excited}} - E_0 > 0$) can generically posses short-range or long-range entanglements.}

{{\bf Short-range entangle states, short-range entanglements (SRE) and SPT states}} are defined as those gapped quantum \emph{ground states} which can be deformed via 
local unitary transformations (LUT) to a trivial tensor product state once we remove, \emph{part of} or \emph{all of}, the internal global symmetries \cite{CGL1314}. Namely, 
along the deformations to a trivial tensor product state, the LUT may break some internal global symmetry of the state.
Gapped SRE states are also named to be \emph{SPT states}.

{{\bf Long-range entangle states, long-range entanglements (LRE) and topological orders}}
are defined as those gapped quantum \emph{ground states} which \emph{cannot} be deformed via 
local unitary transformations (LUT) to a trivial tensor product state, even if we remove all internal global symmetries.
Gapped LRE states are also named to be \emph{topological orders}.

{By this definition \ref{gapp-EE}, we can also rephrase \ref{anom-free} as
 a well-defined quantum field theory (in \ref{well-QFT}) is {all lattice obstruction-free} 
 (\ref{anom-free}) if it can be realized as the 
low energy effective boundary theory of a gapped bulk lattice system 
whose bulk has no LRE (\ie no topological order) and no SRE (\ie no SPT state),
thus the bulk has no entanglement structure at all as a gapped trivial tensor product state.
}
{Readers should be cautious that although this gapped bulk alone has \emph{no entanglement},
the boundary theory (such as an all anomaly-free QFT)
can be \emph{highly-entangled} and can have \emph{gapless states}.}

\item \label{anom-free-2} \textbf{All anomaly-free} (\ie here free of all invertible bosonic and fermionic b-anomalies):

The recent development suggests that all anomaly-free conditions of $d+1$D $G$-symmetric QFT can be understood as the
QFT can live on the boundary of a trivial cobordism class of a trivial invertible topological quantum field theories (iTQFT) 
from a corresponding cobordism group \cite{FH160406527} or its higher-symmetry and higher-classifying space generalization \cite{W2,HAHSII1912.13504,HAHSIII1912.13514}: 
\bea \label{eq:Omegad+2}
\Omega^{d+2}_G.
\eea
Namely, the trivial iTQFT is the trivial element 0 in the $\Omega^{d+2}_G$.
Let us explain this development below. 

't Hooft anomaly is a property that the global symmetry of the theory cannot be
made onsite on a lattice, thus there is an obstruction to gauge the
non-onsite symmetry, which is called the anomalous symmetry
(\ref{m1}) \cite{W1313,Wang:2017loc}. 
Dynamical gauge anomaly is a property that its theory is ill-defined (discussed in \ref{m3}).  How to classify the property of non-onsite global
symmetries or seemly ill-defined theories?  

The previous anomaly inflow picture relates the anomalous
\emph{non-interacting field theories} or \emph{non-interacting lattice models} (\ref{localLatt}) to the boundary of one higher
dimensional bulk \cite{CH8527,F0634}. \Ref{W8597} systematically described anomalies in
field theories in terms of topological invariants in one higher dimension
(such as the index of a 
Dirac operator),
which turn out to be cobordism invariants \cite{DFh9405012}. However, to
construct an \emph{interacting}  lattice regularization of a field theory, we
need to classify anomalies in \emph{interacting} field theories and \emph{interacting} lattice models.  
\Ref{W1313} attempts to classify anomalies in
interacting lattice models, via topological orders and symmetry-protected
topological states (SPTs) of interacting lattice models in one higher dimension.

\end{enumerate}

Let us introduce a few different concepts of anomalies  as the definitions of terminology:

\begin{enumerate}[label=\textcolor{blue}{Def.~\arabic*}:, ref={Def.~\arabic*}]

\item \textbf{b-anomaly} ($\equiv$ boundary {defined} anomaly): \label{b-anom} \\
There are anomalous theories that can be realized as the low energy effective
boundary theory of a gapped local interacting \emph{lattice} model in one
higher dimension, where the global symmetry, if any, is realized as an onsite
symmetry for the whole bulk-boundary coupled system.  However, the effective
symmetry, if any, on the effective boundary theory alone is non-onsite.
There is an obstruction to gauge the non-onsite symmetry \cite{W1313,
Wang:2017loc}, because the {standard} gauging only works for an onsite symmetry:  
{Because there is no canonical way to input the gauge variables on the links between ``non-local sites'' where the non-onsite symmetry acts.}
The obstruction of gauging is the same phenomenon happened in 't Hooft anomalies.
We will call this kind of anomalies as the \emph{b-anomalies}, which include the 't
Hooft anomalies {(associated with some internal global symmetry)}, gravitational anomalies 
{(associated with no internal global symmetry),} 
and {their mixed anomalies}.  

\item \textbf{r-anomalies} ($\equiv$ radical anomaly): \label{r-anom}\\ There are also
anomalous theories that cannot be realized as the low energy effective boundary
theory of any gapped local lattice model in one higher dimension.  We will call
this kind of anomalies as the \emph{r-anomalies}, which include the dynamical gauge
anomalies.  A theory with an r-anomaly is simply an ill-defined quantum theory.

However,  for a dynamical gauge theory with an r-anomaly, very often, we
un-gauge the theory to turn the dynamical gauge field (on the link or on $n$-simplex) into a
global symmetry transformation (onsite or on $(n-1)$-simplex). The resulting un-gauged quantum theory may have a
b-anomaly ('t Hooft anomaly) instead of an r-anomaly. (See examples below.)

\item \textbf{invertible v.s. non-invertible anomalies}: \label{invertible-anom}\\
There are invertible anomalies that can be canceled by other anomalies. (The
anomalies discussed in the field theory literature are mostly invertible
anomalies.)  {Invertible anomalies form an abelian group, such as an infinite
integer group $\Z$ (\ie a perturbative local anomaly, captured by a Feynman
diagram loop calculation) or a finite group $\Z_n$ of some positive integer $n$
(\ie a non-perturbative global anomaly), or the product groups of $\Z$ and
$\Z_n$.  The invertible anomaly labeled by an abelian group element $g$ can be
canceled by an inverted anomaly labeled by an inverted abelian group element
$g^{-1}$.} There are also non-invertible
anomalies\cite{W1313,KW1458,KZ150201690,FV14095723,M14107442,JW190513279} that
cannot be canceled by any other anomalies.  

\item \textbf{bosonic v.s. fermionic anomalies}: \label{bosonic-fermionic-anom}\\
There are bosonic anomalies where the local operators in
the corresponding anomalous theories are all bosonic \cite{Wang2014tia1403.5256,KapustinThorngren2014zva1404.3230}. 
There are fermionic
anomalies where some local operators in the corresponding anomalous theories
are fermionic. {For example, a Spin(10) chiral Weyl fermion theory has an internal 
symmetry Spin(10) $\supset \Z_2^f$ containing the fermion parity,
thus we will need to classify possible 
\emph{fermionic anomalies} of the interacting fermionic theory (later in \eqn{eq:SpinSpinN}) in order to
classify all of its anomalies.\footnote{However, once the [Spin(10)] $\supset \Z_2^f$ is gauged thus 
the fermion parity $\Z_2^f$ is gauged in the Spin(10) chiral Weyl fermion theory,
it becomes a Spin(10) chiral gauge theory, where all local gauge-invariant operators are bosons.}}
\end{enumerate}

{For more examples},
\begin{itemize}
\item
A 1+1D chiral complex Weyl fermion theory with a Hamiltonian, 
$H = \ii \psi^\dag \prt_x \psi$ and a 1-component complex Weyl spinor $\psi$, has 
$$
\text{a fermionic invertible b-anomaly.}
$$  
It is invertible because the anomaly has a $\Z$ class as a group classification.

\item {A 3+1D Weyl fermion doublet coupled to a probed (thus non-dynamical) $\SU(2)$ background gauge field 
has the Witten $\SU(2)$ anomaly \cite{W8224} as a type of 't Hooft anomaly of the SU(2) global symmetry, 
which is 
$$
\text{
a fermionic invertible b-anomaly.}
$$ 
It is fermionic because the SU(2) $\supset \Z_2^f$ has the fermion parity 
at its $\Z_2$ center. It is invertible because the anomaly has a $\Z_2$  class as a group classification.}

\item A 3+1D Weyl fermion doublet coupled to a dynamical $\SU(2)$ gauge field
has the Witten $\SU(2)$ anomaly \cite{W8224}, which is 
$$
\text{a bosonic invertible
r-anomaly.}
$$  It is bosonic since {all the local operators are gauge invariant
and bosonic}.  Namely, the SU(2) ($\supset \Z_2^f$) is dynamically gauged, thus
the fermion parity $\Z_2^f$ is also gauged and the full theory is bosonic. 
It is an r-anomaly, since the Weyl fermion coupled to this SU(2) gauge theory cannot
be realized as a boundary of any gapped local bosonic lattice model \cite{WWW}. 
However, if we un-gauge the SU(2) of this ill-defined gauge theory, then its
bosonic invertible r-anomaly (\ref{r-anom}) becomes the previous fermionic invertible b-anomaly.

\item A $\Z_2$ gauge theory in 2+1D or above with only $\Z_2$ charge
excitations has 
$$
\text{a bosonic non-invertible b-anomaly,}
$$ 
realized as a boundary
theory of a one-higher-dimensional $\Z_2$ gauge theory which is a topological
quantum field theory (TQFT).  \end{itemize}

The classification in \Ref{W1313} is a classification of all b-anomalies in
terms of the topological orders \cite{W9039} or symmetry-protected topological
(SPT) states \cite{GW0931,CLW1141,CGL1314} in local \emph{lattice models} in
one higher dimension.  A b-anomaly is invertible if it is characterized by an
SPT state or an invertible topological order \cite{K1459,KW1458, WGW1489, F1478}
in one higher dimension.  In this work, we will only focus on 
 the invertible b-anomalies. 

From now on, by \emph{anomalous field theory}, we will specifically mean a
well-defined quantum field theory (\ref{well-QFT}) with at most some invertible
b-anomalies (defined in \ref{b-anom}).  {In this work, we only study 
well-defined \emph{quantum field theories} (\ref{well-QFT})
as the effective low energy theory of the boundary of local lattice models  (\ref{localLatt}).
So, we exclude theories with the r-anomaly
(defined in \ref{r-anom}) since they are not well-defined quantum theories (by the norm of both  \ref{localLatt}  and \ref{b-anom}, and the standard lore).}

According to the above classification, an anomaly-free (\ref{anom-free-2}) well-defined quantum {field theory} (\ref{well-QFT})\footnote{Thus
here the \emph{all anomaly-free} condition 
for 
a well-defined quantum field theory (\ref{well-QFT})
specifically satisfies: 
\begin{itemize}
\item free of b-anomalies in \ref{b-anom},
\item
free of all invertible anomalies in \ref{invertible-anom},
\item free of bosonic and fermionic anomalies in \ref{bosonic-fermionic-anom}.
\end{itemize}
\label{footnote:anomaly-free}
} 
is
nothing but a boundary theory of a gapped trivial state (a tensor
product state) on a one-higher-dimensional \emph{lattice}, 
which means
\emph{all lattice obstruction-free} that can be also regularizable in the same dimension (\ref{anom-free}).

 The generalization of the anomaly inflow to a
lattice model with interactions is crucial to obtain this result, since some of
the key concepts, like the tensor product state and the onsite symmetry,
require a lattice (providing the locality of sites) to define.  

With the above terminology definitions, 
we claim a proposition (Prop.):

\begin{enumerate}[label=\textcolor{blue}{Prop.~\Roman*}:, ref={Prop.~\Roman*}]

\item  \label{propoI}
Any well-defined quantum field theory (\ref{well-QFT}) if that is
$$\text{\emph{all anomaly free} (\ref{anom-free-2})}$$ 
with a list of conditions in the footnote \ref{footnote:anomaly-free},
then it is
$$\text{\emph{all lattice obstruction-free} (\ref{anom-free})}$$
required to be regularizable in the same dimension:
Namely, any well-defined QFT
that is \emph{all} anomaly-free can be realized by a local
\emph{interacting} lattice model in the same dimension, where the global
symmetry is realized as an onsite symmetry (or generalized local on-$n$-simplex symmetries) 
\cite{W1313}.\footnote{See Footnote 
\ref{footnote:on-site-symmetry-on-simplex} for the comment on the local symmetry realizations on the lattice.\\[2mm]
Above we propose that: 
\bea
\text{If $\underset{\text{(\ref{anom-free-2})}}{\text{``all anomaly free''}}$ $\to$
then $\underset{\text{(\ref{anom-free})}}{\text{``all lattice obstruction-free.''}}$ }
\eea
However, some well-defined QFTs (\ref{well-QFT}) can be regularized 
on the boundary of one higher-dimensional lattice model, e.g., even if they have b-anomalies in \ref{b-anom}.\\[2mm]
Thus, there is a subtlety about the converse statement.
Only when we restrict the {``all lattice obstruction-free''} 
 requiring QFT to be regularizable in the same dimension and all symmetries realized strictly locally (\ref{anom-free}), then 
the converse statement is also true: 
\bea
\text{If
$\underset{\text{(\ref{anom-free})}}{\text{``all lattice obstruction-free''}}$   
$\to$
then 
$\underset{\text{(\ref{anom-free-2})}}{\text{``all anomaly free.''}}$}
\eea
%
In this work, when we classify invertible 't Hooft anomalies of global symmetries $G$, 
we use the cobordism group $\Omega^{d+2}_G$ in \Eq{eq:Omegad+2}
whose category of manifolds are only smooth and differentiable manifolds. Therefore, 
{we can apply a known mathematical fact that all those smooth and differentiable manifolds are triangulable manifolds, via the Morse theory.}
Thus the anomalies captured in $\Omega^{d+2}_G$ of \emph{smooth and differentiable} manifolds can be triangulated on a lattice
of \emph{triangulable} manifolds. See more comments on \Sec{sec:Cobordism}.
} 


\end{enumerate}

This result can be used to solve the gauged chiral fermion problem via the mirror
fermion approach \cite{W1301}: 
{Given a $d+1$D gauged chiral fermion
theory with a gauge group $G_f \supset \Z_2^f$, we first un-gauge, and obtain a
$d+1$D chiral fermion theory with an internal global symmetry group $G_f \supset \Z_2^f$. Then, we
find a gapped $d+2$D lattice model with a symmetry $G_f \supset \Z_2^f$ whose
boundary realizes the un-gauged $d+1$D chiral fermion theory (the \ref{m1}).
The symmetry $G_f$ is realized as an onsite symmetry of the $d+2$D lattice
model. Next, we determine if the ground state of the bulk gapped $d+2$D lattice
model has a trivial topological order and a trivial SPT state or not. If the
$d+2$D  ground state indeed has no topological order and no SPT
state (which is a trivial tensor product state by \ref{anom-free}), then the $d+1$D un-gauged chiral fermion theory can be realized as the
low energy effective theory of a $d+1$D local lattice model.  Also, the $d+1$D
gauged chiral fermion theory can be realized as the low energy effective theory
of a $d+1$D local lattice model after gauging the onsite symmetry $G_f$.  }

To show the above claim, we can choose the $d+1$D lattice model to be a slab of
the $d+2$D lattice model with \emph{a finite number of layers} in the extra
dimension.  In such a model, the normal sector (or the chiral fermion sector) lives on
one surface of the slab and the mirror fermion sector lives on the other surface of the
slab.    If the normal sector is free of all anomalies, it implies that the
$d+2$D bulk is actually a trivial gapped phase.  
{If so, the mirror sector 
\emph{can be chosen to be} a symmetric gapped boundary and
can be
fully gapped out without breaking the onsite
symmetry \cite{CLW1141,CGL1314,Wang:2017loc}. A detailed explanation is given
in Section \ref{props}. Since the $d+2$D slab has only finite layers, the $d+2$D
slab is actually a $d+1$D lattice model with \emph{finite orbitals per site}.  Last,
we gauge the onsite symmetry to obtain a gauged chiral fermion theory.  }

Thus, the above understanding suggests:
\begin{enumerate}[label=\textcolor{blue}{Prop.~\Roman*}:, ref={Prop.~\Roman*}]
 \setcounter{enumi}{1}
\item   \label{propoII}
{Any $d+1$D gauged chiral fermion theory} (\ref{m2}), 
{that can be realized as the low energy
effective boundary theory of a $d+2$D gapped local lattice model in one higher
dimension} (\ref{well-QFT}), {can be realized as the low energy effective theory of a local
lattice model in the same $d+1$D dimension (\ref{anom-free}), as long as the theory is free of {all}
anomalies} (given by \ref{anom-free-2} and Footnote \ref{footnote:anomaly-free}). 
\end{enumerate}

We remark that for a certain anomalous $d+1$D chiral fermion theory with
an internal symmetry group $G_f$, their corresponding $d+2$D topological/SPT
orders may have a gapped boundary that does not break the $G_f$ symmetry, but
has a non-trivial $G_f$-symmetric anomalous boundary topological order \cite{VS1306,Wang:2017loc} --- 
the low energy theory of topological order may be a $d+1$D $G_f$-symmetric {topological quantum field theories (TQFT)} canceling
the same 't Hooft anomaly of $d+1$D chiral fermion theory.  

For such an anomalous $d+1$D chiral fermion theory, we can have a lattice model in the
same $d+1$D dimension that exactly realizes all the low energy particles of the
anomalous chiral fermion theories.  However, the full low energy effective
theory of the lattice model will contain an extra gauge field for a
finite gauge group $G_\text{extra}$, prescribing the non-trivial anomalous $d+1$D
topological order and TQFT.  Thus, if we only concern about low energy particles, even
some anomalous gauged chiral fermion theories can be regularized by lattice
models in the same dimension \cite{W1301}. But the lattice models will also
produce an extra $d+1$D $G_\text{extra}$-gauge theory with \emph{no} additional low energy
particles, but may give rise to additional extended objects such as string and brane excitations from the TQFT.  


%

It is well-known that a {Spin(10) chiral fermion theory} (\ref{m1}) is 
free of all \emph{perturbative} 't Hooft anomalies;
similarly, it is also well-known that a $\Spin(10)$ chiral gauge theory (\ref{m3})
is free of all \emph{perturbative} dynamical gauge anomalies \cite{A6926,BJ6947}.
%
%
%
%
%
%
But it is not known before if the {Spin(10) chiral fermion theory} (\ref{m1}) is 
free of all other \emph{non-perturbative global anomalies} (of 't Hooft anomalies) or not.  
Thus, it is also not known in the past literature if the $\Spin(10)$ chiral gauge theory (\ref{m3}) is 
free of all other \emph{non-perturbative global anomalies} (as dynamical gauge anomalies) or not. 

\Ref{W1301} provides an argument that the $\Spin(10)$ chiral fermion theory
is free of all anomalies, by proposing a sufficient condition: \emph{A gauged
chiral fermion theory in a $d+1$-dimensional spacetime with a gauge group $G_f$
is free of all anomalies if} (0) \emph{it can be realized as a low energy effective
boundary theory of a gapped local lattice model in one higher dimension} (\ref{well-QFT}), (1)
\emph{there exists a non-zero Higgs field  that makes all the fermions massive, and}
(2) \emph{$\pi_n(G_f/G_\text{grnd})=0$ for $0\leq n\leq d+2$, where $G_\text{grnd}$
is the unbroken gauge symmetry group for the non-zero Higgs field.} 
The chiral fermions satisfying the above conditions can be gapped out by direct interactions
or boson-induced interactions without breaking the $G_f$ symmetry, even when the
fermion mass term is forbidden by the symmetry.  This new mechanism to give
fermions an effective energy gap (or an effective mass) 
is referred to as ``mass without mass term \cite{BZ150504312}.''
{But the above statement is based on an assumption that a smooth
orientation fluctuation of Higgs field  can give rise to a symmetric disordered
phase.} Some other related approaches have also been {proposed \cite{Wang:2013yta,
YBX1451,YX14124784,BZ150504312, Wang:2018ugf}.}

In this work, we do \emph{not} require the proposed conditions of
\Ref{W1301} above, nor need the assumption of new fluctuating Higgs fields in
\Ref{W1301}.  Instead, we will independently and rigorously show that the above
Spin(10) chiral fermion theory (\ref{m1}) is indeed \emph{free of all 't Hooft
anomalies} by a cobordism group approach (\eqn{eq:SpinSpinN}), {and thus it can
be defined on a 3+1D lattice}; which can become a Spin(10) gauged chiral
fermion theory (\ref{m2}) by coupling to a Spin(10) background gauge field, or
become a $\Spin(10)$ chiral gauge theory (\ref{m3}) by dynamically gauging
Spin(10).

\subsection{Gapped boundary of a state
with a trivial invertible topological order with symmetry
}
\label{props}

{
In the following, 
we show that:\\
(1) There \emph{exists} a 3+1D gapped boundary for the above lattice model without breaking the Spin(10) symmetry at the low energy.\\
(2) There \emph{exist} non-perturbative interactions to gap the mirror world chiral fermions without breaking the Spin(10) symmetry.
}

\begin{figure}[!h]
\centering
(a) \includegraphics[scale=.7]{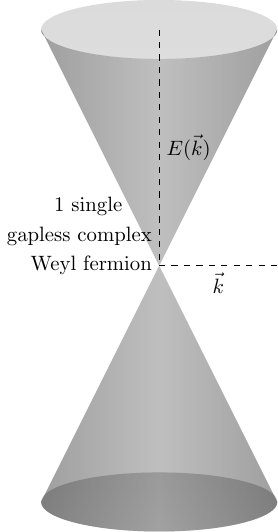}
(b) \includegraphics[scale=.72]{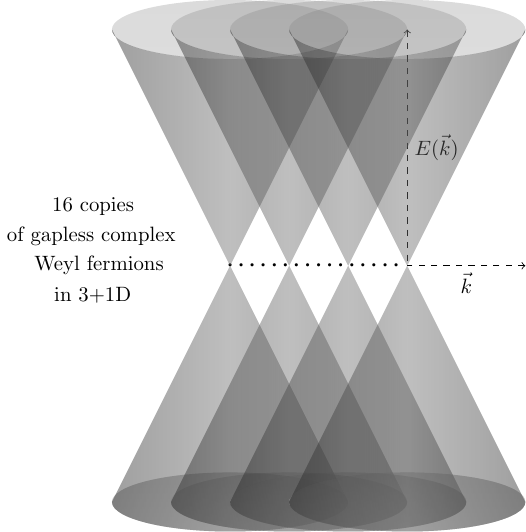}
(c)\includegraphics[scale=.72]{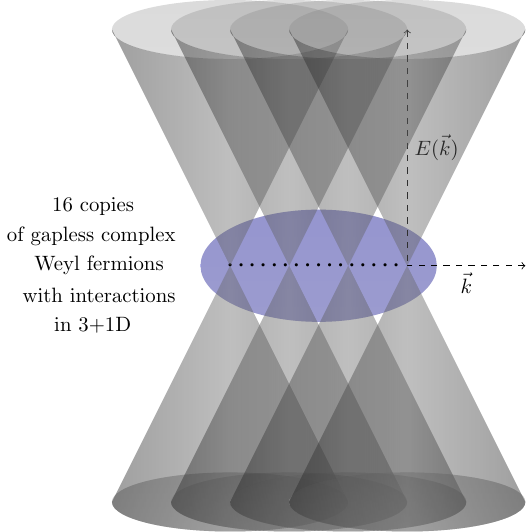}
\caption{\\
(a) A lattice construction of  a single Weyl fermion is given in Section
\ref{so10}, the subfigure shows the gapless energy spectrum 
$E(\vec{k})$ of
Brillouin zone in the schematic 3-dimensional momentum $\vec{k}=({k}_x,{k}_y,{k}_z)$-space with a linear dispersion $|E(\vec{k})| \propto c |\vec{k}|$ for some
effective speed of light $c$. \\
(b) The 16 copies of the same lattice model \eq{H4d} give rise to the 3+1D Weyl fermions at the low energy in the 16-dimensional
spinor representation of the Spin(10) on the lattice boundary shown in Section
\ref{so10}. The 16 gapless Weyl points (schematically the 16 dots $\bullet$) may be separated but can be tuned to the same point on the $\vec{k}$-space Brillouin zone.
We show this Spin(10) chiral Weyl fermion theory is free from all 't Hooft anomalies  via a cobordism theory in \Sec{sec:Cobordism}.\\
(c) There are two ways to obtain the symmetric gapped boundary for the bulk of the 16 copies of the lattice model:
First, via \ref{propoi},
there exists a \emph{symmetric gapped boundary} for the corresponding $d+2$D bulk regularized lattice model (without the need to access from gapping out the gapless theories from interactions. 
Second, via \ref{propoiii},
there exist \emph{non-perturbative symmetric interactions} to fully gap this well-defined all anomaly-free Spin(10) chiral fermion theory with
16 Weyl fermions in 16-dimensional spinor representation of the Spin(10). (Schematic interactions are drawn in the shaded blue region.)
In this work, we only prove \ref{propoi}, but we suggest some supportive evidence for  \ref{propoiii} but without proving  \ref{propoiii}.
However, applying only \ref{propoi} (but without requiring \ref{propoiii})
is sufficient enough for us to construct the Spin(10) chiral fermion theory on the lattice via  \ref{propoI}.
} 
\label{fig:1Weyl}
\end{figure}

The focus of this section is on showing the \emph{existence} (in the
mathematical sense), instead of proving the \emph{constructions} (which may not
be \emph{unique} for the \emph{uniqueness} in the mathematical sense).  In
section \ref{so10}, we provide the 16 copies of the lattice model \eq{H4d} give
rise to the 3+1D Weyl fermions in the 16-dimensional spinor representation of
the Spin(10) on the lattice boundary, see Fig \ref{fig:1Weyl}.

In order to show \ref{propoI} which consequently includes also \ref{propoII},
we break down this proposition into several relatedly helpful sub-propositions. For
any well-defined $d+1$D QFT (defined in \ref{well-QFT}) 
that is \emph{all} anomaly-free (defined in \ref{anom-free-2}) with an internal symmetry $G_f$,
which can live on the boundary of $d+2$D bulk regularized lattice model, 
we aim to show that (which we focus on the spatial dimension $d=3$):
{ 
\begin{enumerate}[label=\textcolor{blue}{Prop.~\roman*}:, ref={Prop.~\roman*}]
\item \label{propoi}
There exists a \emph{symmetric gapped boundary} for the corresponding $d+2$D bulk regularized lattice model. 
This $d+1$D symmetric gapped boundary does not break any internal symmetry $G_f$ of the whole bulk-boundary system, and 
does not contribute any ground state degeneracy (neither symmetry-breaking degeneracy, nor topological degeneracy \cite{Wang2012am1212.4863,Kapustin2013nva1306.4254}).


\item \label{propoiii}
There exist \emph{non-perturbative symmetric interactions} to fully gap this well-defined all anomaly-free $d+1$D QFT, via
deforming the QFT by adding any all anomaly-free gapless or gapped sectors, 
while still preserving the full $G_f$ internal symmetry,
without any symmetry-breaking and without contributing any degeneracy (neither symmetry-breaking degeneracy nor topological degeneracy). 

\end{enumerate}
}

We will see that showing \ref{propoi} is sufficient enough to show that
 \ref{propoI} is also true. In other words, we only need  \ref{propoi} but do not need to prove \ref{propoiii}, 
 in order to prove \ref{propoI}.\\

\noindent
$\bullet$ To show \ref{propoi}, we first note that 
by a symmetric gapped boundary, we also mean that the ground state energy $E_0$ (of this whole bulk-boundary system) to its higher energy excited states (at energy $E_1, \dots$, etc.), 
are separated by a finite energy gap $\Delta_E = E_1 - E_0 > 0$.
Of course, by defining the energy gap $\Delta_E> 0$ here, we should first set-up a toy-model system with only such a $d+1$D symmetric gapped boundary and a fully gapped $d+2$D bulk.
(We either have only this gapped boundary and without other boundaries, or other boundaries are also fully gapped.) 

{
If the gapped d+2D bulk has also a \emph{symmetric gapless} boundary (say on a d+1D boundary A)
other than the \emph{symmetric gapped} boundary  of \ref{propoi} (say on another d+1D boundary B).
Thus the gapless boundary A contributes to the low energy spectrum at the infrared (IR) of a tiny energy sub-gap 
\bea
\delta_{E,\rm{A}} \simeq \exp(-L/\xi)
\eea
which scales exponentially over the linear system size $L$ over the correlation length $\xi$;
the gapped boundary B contributes to the energy spectrum only at the higher energy at a deeper UV of a finite energy gap 
\bea \label{eq:DeltaB}
\Delta_{E,\rm{B}} \simeq \Delta_E >0,
\eea
mentioned earlier.
Then the whole bulk-boundary system would become \emph{gapless} instead of being gapped.
}

{The important issue is that when the $d+2$D gapped bulk has no entanglements (\ie no LRE nor SRE by \ref{gapp-EE}), then 
``the 
d+1D symmetric gapless boundary A'' 
and 
``the 
d+1D 
symmetric gapped boundary B'' actually cannot affect each other, thus are isolated from each other. See more in Appendix
\ref{sec:Energy-mutual-entanglement}.}

{
To proceed showing \ref{propoi}, if the bulk regularized lattice is in the
gapped trivial phase (\ie has a gapped trivial tensor product ground state), 
we can make a boundary by first deforming the bulk ground state (by symmetry-preserving LUT in \ref{gapp-EE}) into a  tensor product state. 
Such a deformation does \emph{not} close the energy gap since the bulk state is already in the symmetric gapped trivial phase. 
The trivial tensor product state \emph{always} can have a gapped boundary respect to a 
trivial vacuum\footnote{The trivial tensor product state \emph{always} can have a gapped boundary respect to a trivial vacuum,
because the trivial tensor product state is itself the same phase indistinguishable as the trivial vacuum.
Thus its gapped boundary simply is the trivial gapped domain wall between the same phase \cite{KitaevKong2012.1104.5047, Lan2014uaaLanWangWen1408.6514}. 
} 
--- by saying so,
we mean that we set the energy scale of the trivial vacuum (normally to below some energy scale such as a finite energy $\Delta_E >0$, or below an infinite energy gap $\Delta_E \to \infty$)
to be the same as the energy scale of the gapped boundary (say on B) $\Delta_E
\simeq \Delta_{E,\rm{B}}  >0$ in \eqn{eq:DeltaB}.
We note that the above deformation respects the onsite symmetry (if
any), and the resulting tensor product state also respects the onsite symmetry.
The gapped boundary does not break the onsite symmetry, thus 
has no  \emph{symmetry-breaking degeneracy}. 
Since the symmetric gapped boundary has no entanglements, thus 
has no   \emph{topological degeneracy} (because topological degeneracy \cite{Wang2012am1212.4863,Kapustin2013nva1306.4254} are due to LRE defined in \ref{gapp-EE}). 
}\\

The above completes our proof of \ref{propoi}.\\

In Appendix \ref{sec:app-symmetric-gapped-boundary}, we can also show 
\ref{propoi} by a second viewpoint:
a derivation from
the \emph{classification of quantum phases of matter and their phase transitions}.

This second viewpoint from the \emph{classification of quantum phases of matter} 
 shows that there is \emph{no need for an energy-gap closing phase transition}.
 By maintaining a finite energy gap $\Delta_E$ between two phases, there must exist a \emph{symmetric gapped boundary} between two phases, 
thus we have given an alternative proof of \ref{propoi}.

The slight conceptual difference between the first viewpoint and the second viewpoint is that, 
the first is about the \emph{one-spatial-dimensional-lower phase boundary} in $d+1$D between two $d+2$D phases,
while the second is about no need for the \emph{phase transition} in $d+2$D between two $d+2$D phases in a quantum phase diagram 
(at zero temperature $T=0$) by tuning a certain coupling $g$.\\

\noindent $\bullet$  {To show \ref{propoI},
we consider a $d+2$D bulk regularized lattice model that
realizes the $d+1$D QFT as its boundary theory by \ref{well-QFT}.  
We choose the bulk lattice model to be
a slab of finite thickness, such that one boundary of the slab realizes the
QFT (Boundary A), and the other boundary is a symmetric gapped boundary (Boundary B) in \ref{propoi}.
We apply the \ref{propoi} proven earlier. 
Here a slab of finite thickness is always achievable for this system (especially for the gauged chiral fermion problem of \ref{m1} and \ref{m2}), 
because of the isolation between two $d+1$D boundaries A and B due to
Appendix \ref{sec:Energy-mutual-entanglement}'s Remark \ref{i-E-scale} on the isolation of the energy scale and 
Remark \ref{ii-EE} on the isolation of the mutual entanglement,
see Appendix \ref{sec:Energy-mutual-entanglement} on
the {energy scale and mutual entanglement between gapless and gapped boundaries}.}

Thus the low energy physics of the $d+2$D slab is described by this $d+1$D QFT.  A lattice model of this $d+2$D slab
of a finite thickness can be constructed explicitly as a lattice model in one lower dimension ($d+1$D),
by rewriting the ``quantum Hilbert space associated with different lattice sites along the finite width thickness w (\ie an extra small dimension along w)''
to ``quantum Hilbert space associated with finite orbitals per site'' in $d+1$D.\\

This completes our proof of \ref{propoI}.\footnote{As we said earlier, we do not need \ref{propoiii} to show \ref{propoI}.} 
%
\onecolumngrid
%
\subsection{A deformation class of all anomaly-free well-defined QFTs}

\twocolumngrid

\noindent $\bullet$  {For \ref{propoiii},  we again consider a bulk regularized lattice model
that realizes the QFT as its boundary theory by \ref{well-QFT}.  Since the same bulk model can
also have a symmetric gapped boundary according to \ref{propoi}, thus we ask: How to modify the
symmetric interactions in the QFT to make it into a fully symmetric gapped theory describing the
symmetric gapped boundary?}

One key ingredient is that there are more degrees of freedom given by the full Hilbert space 
to help us reach the goal. We may be able to access the symmetric gapped phase not only within this specific well-defined  \emph{all anomaly-free} QFT,
but also higher energy spectrum by engineering all possible degrees of freedom and their symmetry-preserving local interactions.


We do not have a direct proof of \ref{propoiii}, but we have several supportive evidence to argue \ref{propoiii} should be true:\\

\noindent
($\alpha$)  
``\emph{The deformation classes of QFTs}'' advocated by Seiberg \cite{NSeiberg-Strings-2019-talk}:
Given a continuum QFT with some energy scale $\Lambda$, with a given \emph{global symmetry} and derivable \emph{'t Hooft anomaly} of global symemtry.
We are allowed to add arbitrary degrees of freedom and new fields preserving 
the symmetries (and selection rules) and with no additional anomaly (without modifying the original {'t Hooft anomaly})
at some energies. 
The new degrees of freedom \emph{do not} directly affect the dynamics at lower energies. 
Next, we can deform the parameters of this larger theory with the new degrees of freedom, by making the new degrees of freedom interacting with
the original QFT, which \emph{do} affect the dynamics. This is a much larger space of theories, which can land into different new phases with different dynamics. 
Seiberg names all these possible deformations of QFT  as a \emph{deformation class of the QFT}.

\noindent
Seiberg \cite{NSeiberg-Strings-2019-talk} conjectured that given two QFTs, say partition functions ${\bf Z}_1$ and ${\bf Z}_2$, in the same spacetime dimension with:\\
($i$) the same global symmetry (and selection rules),\\
($ii$)   the same 't Hooft anomalies,\\
we can always add new degrees of freedom at short distances so that we can interpolate between two QFTs:
The two QFTs, ${\bf Z}_1$ and ${\bf Z}_2$, are in the same \emph{deformation class of QFT}.
In other words, this also means that the deformation class of QFT can be determined and defined by the symmetries and the 't Hooft anomalies of QFT.
Seiberg's conjecture is in fact shown to be true for many examples. \\

\noindent
What we claim on \ref{propoiii} is indeed a special case of Seiberg's proposal  \cite{NSeiberg-Strings-2019-talk}:
We consider the \emph{deformation class of the anomaly-free well-defined QFT}, containing
the trivial gapped phase (e.g. a symmetric gapped Spin(10)  boundary) and a gapless phase (e.g. a symmetric gapless Spin(10) chiral fermion theory), both 
have ($i$)  
 the same global symmetry (and selection rules)
and ($ii$)  no 't Hooft anomalies (we will show via a cobordism theory in \Sec{sec:Cobordism}).
If  Seiberg's proposal  \cite{NSeiberg-Strings-2019-talk} is true, our proposal must also be true.\\

\noindent
($\bt$) We can start from the $d+1$D symmetric \emph{gapless all-anomaly-free} theory, 
and adding new $d+1$D symmetric \emph{gapped all-anomaly-free} sectors (this is analogous to Seiberg's proposal \cite{NSeiberg-Strings-2019-talk}). \\

\noindent
Moreover, 
we can also add additional \emph{gapless all-anomaly-free} sectors in the  \emph{various possible representations} (Rep) of symmetry.\footnote{For example, the trivial Rep 
of pairs of left and right moving 3+1D Weyl fermion $\psi_L$ and $\psi_R$ 
in the trivial Rep of Spin(10). Then adding their mass term, e.g. $m (\psi_L^\dagger \psi_R+ \psi_R^\dagger\psi_L)$
(only for these additional trivial Rep gapless sectors) do not break the Spin(10) symmetry.}
The symmetry organizes the (both gapped and gapless) energy eigenstates in the energy spectrum into \emph{various possible representations} of the symmetry group,
whose selection rules constrain the interactions and dynamics between states in the energy spectrum.\\

\noindent
Based on ($\alpha$) and ($\bt$), we propose that \ref{propoiii} is also true.\\

\section{
Construct a local bosonic lattice model
 realizing a 3+1D $\Spin(10)$ gauged chiral fermion theory}

\label{bLat}

Below we will use the slave-particle/parton
approach \cite{BA8880,AZH8845,DFM8826} to explicitly construct a 4+1D \emph{local
bosonic} lattice model, whose boundary can give rise to the dynamical
$\Spin(10)$ chiral gauge theory coupled to Weyl fermions (\ref{m3}) in the 16-dimensional
representation.  We start with a fermionic model on a 4D cubic lattice. On each
site we have $10 N_f$ complex 
fermions $\hat \psi_{\al,m}$, $\al=1,\cdots,10$,
$m=1,\cdots,N_f$.  So on each side, there are $2^{16N_f}$ states.  Now we
project into the even fermion subspace on each site, and turn the fermionic
model into a bosonic model with $2^{16N_f-1}$ states per site.  The Hamiltonian
for such a bosonic model is given by
\begin{align}
 \hat H_1 &=\sum_{\<ij\>} \sum_{\al\bt} (\hat \chi_{ij}^{\al\bt})^\dag \hat \chi_{ij}^{\al\bt}
+ \sum_i (-)^i \hat n_i,
\\
\hat \chi_{ij}^{\al\bt} &= \sum_m \hat \psi_{\al,m,i}^\dag \hat \psi_{\bt,m,j} 
,\ \ \ \
\hat n_{i} = \sum_{m,\al} \hat \psi_{\al,m,i}^\dag \hat \psi_{\al,m,i}. 
\nonumber 
\end{align}
The above model has $[\U(16)]^{N_\text{site}}$ local symmetry.  In the large
$N_f$ limit, $\hat \chi_{ij}^{\al\bt}$ is weakly fluctuating
and can be replaced by  $\chi (\ee^{\ii A_{ij}})_{\al\bt} =\<\hat \chi_{ij}^{\al\bt}\> $ expectation value and $A_{ij}$ is a $16\times 16$ Hermitian matrix to
describe the $\U(16)$ gauge fluctuation.
This leads to
the following emergent $\U(16)$ gauge theory (at a mean-field level)
\begin{align}
 \hat H_1^\text{mean} &=\sum_{\<ij\>} \sum_{\al\bt,m} \Big[
\hat \psi_{\al,m,i}^\dag \chi^* (\ee^{-\ii A_{ij}})_{\al\bt}\hat \psi_{\bt,m,j}
+h.c.\Big]
\nonumber\\
&\ \ \ \ \
+ \sum_i (-)^i \hat n_i .
\end{align}
The ground state is given by
$A_{ij}=0$. The emergent fermions are in a fully gapped product state and the
bosonic model $\hat H_1$ gives rise to a $\U(16)$ gauge theory at low energies.

Next, we reduce the $\U(16)$ gauge theory to the $\Spin(10)$ gauge theory by adding
a term
\begin{align}
 \hat H_2 = \sum_{i,m} \Ga^{\al\bt\ga\la}
\hat \psi_{\al,m,i}
\hat \psi_{\bt,m,i}
\hat \psi_{\ga,m,i}
\hat \psi_{\la,m,i} + h.c.
\end{align}
to break the $[\U(16)]^{N_\text{site}}$ local symmetry to a
$[\Spin(10)]^{N_\text{site}}$ local symmetry where the fermions
$\psi_{\al,m,i}$ form the 16-dimensional spinor representation.  
$\Ga^{\al\bt\ga\la}$  is the antisymmetric tensor which is invariant under the
$\Spin(10)$ transformations.  The 4+1D bosonic model $\hat H_1 + \hat H_2$ will
give rise to an emergent $\Spin(10)$ gauge theory with the fermions in a fully
gapped product state.  Those fermions are also gapped on the boundary.

To have gapless Weyl fermions on the boundary,
we add the third term
\begin{align}
 \hat H_3=\sum_{\v i\v j} (t_{\v i\v j}^{ab} \hat c^\dag_{\al,a,\v i} \hat c_{\bt,b,\v j}\hat \chi_{ij}^{\al\bt} +h.c.).
\end{align}
Now on each site, we have fermions $\hat \psi_{\al,m}$ and $\hat c_{\al,a}$, but
we still project into the subspace with even fermion per site. The $ \hat H_1
+\hat H_2 +\hat H_3$ acts within this subspace.  So the model is still a
lattice bosonic model.  When $t_{\v i\v j}$ is given by \eqn{H4d}, the model $
\hat H_1 +\hat H_2 +\hat H_3$ will give rise to emergent massless Weyl fermions
on the boundary coupled to the $\Spin(10)$ gauge field.

Consider a 4+1D slab of the local bosonic lattice model described by $ \hat H_1
+\hat H_2 +\hat H_3$.  In the main text, based on the complete classification
of 't Hooft anomaly of the group $G={{(\Spin(5) \times \Spin(10))}/{\Z_2^f}}$
(in \eqn{eq:SpinSpinN} and \eqn{topinv}), we have shown that there exists a
boundary interaction, that allows us to gap out the boundary  Weyl fermions
(without inducing a boundary 3+1D topological order \cite{Wang:2017loc}) on one
of the surfaces of the slab.  In this case, the  4+1D slab, with a
\emph{finite} width in the extra dimension, indeed becomes a 3+1D \emph{local
bosonic} lattice model that realizes a 3+1D dynamical $\Spin(10)$ gauge theory
coupled to Weyl fermions in the 16-dimensional spinor representation.


\section{Energy scale and mutual entanglement between gapless and gapped boundaries}
\label{sec:Energy-mutual-entanglement}

{
The important issue is that when the $d+2$D gapped bulk has no entanglements (\ie no LRE nor SRE by \ref{gapp-EE}), then 
``the 
d+1D symmetric gapless boundary A'' 
and 
``the 
d+1D 
symmetric gapped boundary B'' actually cannot affect each other, thus are isolated from each other, in the sense that: 
\begin{enumerate}[label=\textcolor{blue}{(\roman*)}:, ref={(\roman*)}]
\item \label{i-E-scale}
{\bf Energy scale}: Boundary A and Boundary B are \emph{decoupled} below the energy scale $\simeq \Delta_{E,\rm{B}}$. 
But when the energy is above the scale $\gtrsim \Delta_{E,\rm{B}}$, the energy spectra of A, B, and the bulk may affect and mix with  each other together.\\
\item \label{ii-EE}
{\bf Mutual entanglement}:   Although the d+1D symmetric gapless boundary A is highly-entangled (due to the low-lying massless chiral fermions as the energy gapless spectrum),
and the symmetric gapped boundary B is trivially gapped with no entanglements as a tensor product state on Boundary B.
Thanks to the trivial gapped bulk, Boundary A and Boundary B on two sides have no entanglements in between. 
More precisely, if we choose a $d+1$D bipartite cut inside the $d+2$D gapped bulk,
we get a zero bipartite Von Neumann entanglement entropy 
\bea \label{eq:EE}
S_{\text{EE}}={S}(\rho_A)=  -\operatorname{Tr}[\rho_A\operatorname{log}\rho_A] \nn\\
=  -\operatorname{Tr}[\rho_B\operatorname{log}\rho_B] = {S}(\rho_B)=0
\eea
for the mutual entanglement between two sides.
(This understanding is consistent with the entanglement structure discussed in  \ref{gapp-EE}.)
\end{enumerate}
}

For the \emph{lattice regularization of the gauged chiral fermion problem}, we should emphasize that our statements in \ref{i-E-scale} and \ref{ii-EE}
apply to \ref{m1} (a chiral fermion theory) and \ref{m2} (a gauged chiral fermion theory). However, 
we do not intend to apply our statements in \ref{i-E-scale} and \ref{ii-EE} to \ref{m3} (a chiral gauge theory) ---
once we dynamically gauge the internal global symmetry for the bulk-boundary coupled system,
then ``the bulk, Boundary A and Boundary B'' form altogether highly-entangle quantum states (as a dynamical gauge theory).
The $S_{\text{EE}}$ of the dynamically gauged system (\ref{m3}), based on the previous bipartite cut in \eqn{eq:EE}, is generically non-zero.

\section{Show the existence of {symmetric gapped boundary via quantum phase transitions}}
\label{sec:app-symmetric-gapped-boundary}

{A derivation can also be obtained from
the \emph{classification of quantum phases of matter and their phase transitions}. 
To give a proof of \ref{propoi}, all we need to show is that there exists a LUT deformation path (\ref{gapp-EE}) between two bulk gapped quantum phases:\\
(1) Bulk phase: The $d+2$D bulk regularized lattice model which has a symmetric gapped trivial tensor product ground state, with a finite energy gap $\Delta_E$.\\
(2) Trivial gapped vacuum phase (mentioned above).\\
Such that this LUT deformation path satisfies the criteria:\\ 
($\emph{1}$) does not close the energy gaps between Bulk Phase and Trivial Vacuum Phase (\ie no gap closing, thus 
no gapless modes and no zero-mode degeneracy [=ground state degeneracy]). \\
($\emph{2}$) does not break the internal global symmetry given by the Bulk Phase. (\ie $G_f=\Spin(10)$ for the Spin(10) gauged chiral fermion problem.)\\
}
This LUT deformation path can be regarded as a path labeled by $g$ in the quantum phase diagram (at zero temperature $T=0$)  
by tuning a parameter (\ie a coupling constant) $g$ of the lattice Hamiltonian $\hat{H}(g)$
such that the ground state $|\Psi_{\text{g.s.}}(g) \rangle$ is unitarily evolving under this LUT along the deformation path.
Then we can prove the claim of ($\emph{1}$) and ($\emph{2}$), either by ``\emph{proof by a contradiction},'' or by directly ``\emph{constructing such a LUT deformation path}.''

``\emph{A proof by a contradiction}'': Suppose, given by engineering arbitrary symmetry-preserving (e.g. $G_f$) local interactions for the lattice Hamiltonian, 
such a path in the phase diagram is still impossible between two phases (the bulk phase and the trivial gapped vacuum phase).
Then there must be a phase transition between two phases, and the two phases should be different quantum phases --- in fact, they should be \emph{different} SPT phases within the $G_f$ symmetry.
But as we emphasize that both two phases have symmetric gapped trivial tensor product ground states, they must be in the \emph{same trivial} SPT phase, thus the same 
trivial gapped vacuum, at least below the energy gap $\Delta_E$ of the bulk phase. This leads to a contradiction, thus we end the proof successfully.

``\emph{Constructing such a LUT deformation path}'': This path construction is basically what we had in the earlier proof. Since both phases are symmetric gapped
trivial phases (both a trivial SPT phase and a trivial gapped vacuum respect to the $G_f$ symmetry), the LUT deformation path
is simply the deformation to make both symmetric gapped
trivial phases become \emph{exactly the same symmetric gapped trivial tensor product states} in a certain ``\emph{canonical basis}'' respect the $G_f$ symmetry.
(Normally it is known as the \emph{symmetric disordered} phase, where the canonical basis is chosen to be the \emph{dual} variable of the symmetry breaking basis.)

\section{Cobordism theory and
classification of all possible 
invertible anomalies related to SU(5) and SO(10) Grand Unifications}
\label{Appendix:Cobordism}

Here we provide the cobordism group calculations classifying all potential invertible 't Hooft anomalies of SU(5), Spin(10) and Spin(18) chiral fermion theories.
Our calculations are crucial for showing all gauge anomaly free conditions for SU(5), SO(10) and SO(18) Grand Unifications.
Notice that other related work \cite{Garcia-Etxebarria:2018ajm} computes 
$\Omega^{\Spin \times \SU(5)}_D$ and $\Omega^{{\Spin \times \Spin(10)}}_D$
based on a different method, Atiyah-Hirzebruch spectral sequence (AHSS), 
while our work focus on $\Omega^{\Spin \times \SU(5)}_D$ and $\Omega^{{\Spin \times \Spin(10)}/{\Z_2^f}}_D$,
also based on a more powerful Adams spectral sequence.

\subsection{Adams spectral sequence}

The Adams spectral sequence shows:
\bea \label{eq:Adams}
\Ext_{\A_p}^{s,t}(\H^*(Y,\Z_p),\Z_p) \; \Rightarrow \; \pi_{t-s}(Y)_p^{\wedge}, 
\eea
where Ext denotes the extension functor, 
$\A_p$ is the mod $p$ Steenrod algebra, and $Y$ is any spectrum.  
{The $\H^*(Y,\Z_p)$ is an ${\A_p}$-module whose internal degree $t$ is given by the $*$.}
The $\pi_{t-s}(Y)_p^{\wedge}$ is the $p$-completion of 
the ${(t-s)}$-th homotopy group of the spectrum $Y$.
We note that, 
for any finitely generated abelian group $\mathcal{G}$,
then $\mathcal{G}_p^{\wedge}=\lim_{n\to\infty}\mathcal{G}/p^n \mathcal{G}$ is the $p$-completion of $\mathcal{G}$;
if $\mathcal{G}$ contains an infinite group $\Z$, then the $\mathcal{G}_p^{\wedge}$ is the ring of $p$-adic integers.
Here the $\mathcal{G}$ is meant to be substituted by a homotopy group $\pi_{t-s}(Y)_p^{\wedge}$ in \Eq{eq:Adams}.
Here are some explanations and inputs:
\begin{enumerate}[leftmargin=3mm, label=\textcolor{blue}{\arabic*}., ref={\arabic*}] 
\item
{Here the double-arrow 
``$\Rightarrow$'' means ``convergent to.'' 
The $E_2$ page contains groups $\Ext^{s,t}$ 
with double indices $(s,t)$, we reindex the bidegree by $(t-s,s)$. 
There are differentials $d_2$ in $E_2$ page which are arrows from $(t-s,s)$ to $(t-s-1,s+2)$. 
That is, $\Ext^{s,t}\to \Ext^{s+2,t+1}$. Take Ker$d_2$/Im$d_2$ at each $(t-s,s)$, then we get the $E_3$ page. 
Repeat this procedure, we get $E_4$ page, $E_5$ page and so on. 
Finally $E_r$ page equals $E_{r+1}$ page (there are no differentials) for $r \geq N$, we call this $E_N$ page as the $E_{\infty}$ page, we can read the result $\pi_D$ at $D=t-s$.
See further details discussed in \Ref{W2}'s Sec.~2.3.}
\item
{In Adams spectral sequence, we consider $\Ext_R^{s,t}(L,\Z_p)$. Here we have the ring or the algebra $R=\A_p$ or $\A_2(1)$ for $p=2$, and {the $L$ is a $R$-module}. 
The $\A_2(1)$ is the subalgebra of $\A_2$ generated by the Steenrod square $\Sq^1$ and $\Sq^2$.
The index $s$ refers to the degree of resolution, and the index $t$ is the internal degree of the $R$-module $L$. 
Ext groups are  defined by firstly taking a projective $R$-resolution $P_{\bullet}$ of $L$, then secondly computing the (co)homology group of the (co)chain complex $\Hom(P_{\bullet},\Z_p)$.
{A $P_{\bullet}$ is a resolution, which is an exact sequence of modules.}
Here a projective $R$-resolution $P_{\bullet}$ is an exact sequence of $R$-modules $\cdots\to P_s\to P_{s-1}\to \cdots\to P_0\to L$ where $P_s$ is projective for $s\ge0$.
}
\end{enumerate}

\onecolumngrid

\subsection{Thom-Madsen-Tillmann spectrum and Pontryagin-Thom isomorphism}
For $Y=MTG$, where $MTG$ is the Thom-Madsen-Tillmann spectrum $MTG$ 
 of a group $G$,
the Adams spectral sequence shows:
\bea \label{eq:Adams}
\Ext_{\A_p}^{s,t}(\H^*(MTG,\Z_p),\Z_p) \; \Rightarrow \; \pi_{t-s}(MTG)_p^{\wedge}
{\; = \;(\Omega_{D=t-s}^G)_p^{\wedge}}. 
\eea
The last equality is by the generalized Pontryagin-Thom isomorphism, we have an equality between the $D$-th bordism group of $G$
given by $\Omega_D^G$ and the $D$-th homotopy group of $MTG$ given by $\pi_D(MTG)$, namely
\bea  \label{eq:Pontryagin-Thom}
\Omega_D^G=\pi_D(MTG).
\eea

We also compute the cobordism group of topological phases (TP) defined in \cite{FH160406527} as 
\bea
\TP_D(G).
\eea 
The $\TP_D(G)$ classifies deformation classes of reflection positive invertible $d$-dimensional extended topological field theories with symmetry group $G_D$. The $\TP_D(G)$ and the bordism group $\Omega_D^G$ are related by a short exact sequence
\bea \label{eq:Ext-extension}
0\to\Ext^1(\Omega_D^G,\Z)\to\TP_D(G)\to\Hom(\Omega_{D+1}^G,\Z)\to 0.
\eea

{
We can compute the $E_2$ page of $\A_2(1)$-module based on 
Lemma 11 of \cite{W2}.
More precisely, in order to compute $\Ext_{\A_2(1)}^{s,t}(L_2,\Z_2)$, we find a short exact sequence of 
$\A_2(1)$-modules 
\bea  \label{eq:L-extension}
0\to L_1\to L_2\to L_3\to 0,
\eea
then we apply Lemma 11 of \cite{W2} to compute $\Ext_{\A_2(1)}^{s,t}(L_2,\Z_2)$ by the given data of 
$\Ext_{\A_2(1)}^{s,t}(L_1,\Z_2)$ and $\Ext_{\A_2(1)}^{s,t}(L_3,\Z_2)$. 
Our strategy is choosing $L_1$ to be the direct sum of suspensions of $\Z_2$ on which $\Sq^1$ and $\Sq^2$ act trivially, then we take $L_3$ to be the quotient of $L_2$ by $L_1$.
We can use this procedure again and again until $\Ext_{\A_2(1)}^{s,t}(L_3,\Z_2)$ is determined.
}

{
If $G=\Spin\times G'$, then $\B G=\B(\Spin\times G')=\B\Spin\times\B G'$. By definition, the Madsen-Tillmann spectrum $MTG=\text{Thom}(\B G,-V)$ where $V$ is the induced virtual bundle of dimension 0 by the map $\B G\to\B\rm{O}$.
By the properties of Thom space (see the discussions in \Ref{W2}'s Sec.~1.3), we have 
\bea
MT(\Spin\times G')=M\Spin\wedge(\B G')_+.
\eea
The $\wedge$ is the smash product.

Below we will use the \Eq{eq:Adams} and \Eq{eq:Pontryagin-Thom}
to compute the $D$-th bordism group of $G$
given by $\Omega_D^G$.
Then we will use the \Eq{eq:Ext-extension}
and the techniques around \Eq{eq:L-extension}
to compute the $D$-th cobordism group of topological phases of $G$
given by $\TP_D(G)$.

\subsection{Cobordism groups and topological phases for ${\Spin \times \SU(5)}$: SU(5) Grand Unification}


We consider $G=\Spin \times \SU(5)$ for the Georgi-Glashow SU(5) Grand Unification \cite{GG7438}, the Thom-Madsen-Tillmann spectrum $MTG$  of the group $G$ is
\bea
MTG=M\Spin\wedge (\B \SU(5))_+.
\eea
The $T$ in $MTG$ means the $G$-structures are on tangent bundles instead of normal bundles.
For Spin, the Thom-Madsen-Tillmann spectrum $MT\Spin= M\Spin$ is equivalent to the Thom spectrum which splits 
$M\Spin =ko\vee \Sigma^8ko\vee\cdots$. 
The $ko$ is the $(-1)$-connected cover of the real K-theory spectrum.
The $\wedge$ is the smash product and
the $\vee$ is the wedge sum. 
The $(\B \SU(5))_+$ is the disjoint union of the classifying space $\B \SU(5)$ 
and a point.\footnote{For a topological space $X$, it is a standard convention to 
denote that $X_+$ as the disjoint union of $X$ and a point.
Note that the reduced cohomology of $X_+$ is exactly the ordinary cohomology of $X$.}

For the dimension $D=t-s<8$, since there is no odd torsion,\footnote{By computation using the 
mod $p$ Adams spectral sequence for an odd prime $p$, we find there is no odd torsion. \label{ft:no-odd-torsion}} 
by $MTG=M\Spin\wedge X$, then 
the $D$-th homotopy group $\pi_D(MTG)=\pi_D(ko\wedge X)$ for $D<8$.
So for the dimension $D=t-s<8$, we have
\bea\label{eq:ExtA_2(1)}
\Ext_{\A_2(1)}^{s,t}(\H^*(X,\Z_2),\Z_2)\Rightarrow(\Omega_{D=t-s}^G)_2^{\wedge}.
\eea
Hence for $MTG=M\Spin\wedge (\B \SU(5))_+$, 
for the dimension $D=t-s<8$, by \eqref{eq:ExtA_2(1)}, we have the Adams spectral sequence
\bea
\Ext_{\A_2(1)}^{s,t}(\H^*(\B \SU(5),\Z_2),\Z_2)\Rightarrow\Omega_{t-s}^{\Spin \times \SU(5)}.
\eea

The $\A_2(1)$-module structure of $\H^*(\B \SU(5),\Z_2)$ below degree 6 
is shown in \Ref{WW1910.14668}'s Sec.~6's Figure 29, and the $E_2$ page is shown in Figure  \ref{fig:E_2SU5}.
Here we have used the correspondence between $\A_2(1)$-module structure and the $E_2$ page shown in 
Appendix A
of \Ref{WW1910.14668}.

\begin{figure}[!h]
\begin{center}
\includegraphics[scale=.8]{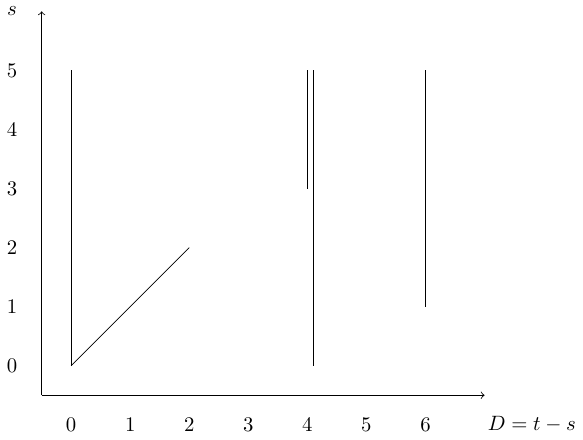}
\end{center}
\caption{Adams chart for $\Omega_D^{\Spin \times \SU(5)}$.}
\label{fig:E_2SU5}
\end{figure}

In Adams chart, the horizontal axis labels the integer degree $D=t-s$ and the vertical axis labels the 
integer degree $s$. 
The differential $d_r^{s,t}:E_r^{s,t}\to E_r^{s+r,t+r-1}$ is an arrow starting at the bidegree $(t-s,s)$ with direction $(-1,r)$.
$E_{r+1}^{s,t}:=\frac{\text{Ker}d_r^{s,t}}{\text{Im}d_r^{s-r,t-r+1}}$ for $r\ge2$. There exists $N$ such that $E_{N+k}=E_N$ stabilized
for $k>0$, we denote the stabilized page $E_{\infty}:=E_N$.

To read the result from the Adams chart in Figure \ref{fig:E_2SU5}, we look at the stabilized
$E_{\infty}$ page, one dot indicates a finite group $\Z_p$, 
a vertical finite line segment connecting $n$ dots indicates a finite group $\Z_{p^n}$. 
But when $n=\infty$, the infinite line connecting infinite dots indicates an infinite group, an integer $\Z$. 
Here $p$ is given by the mod $p$ Steenrod algebra $\A_p$ in \Eq{eq:Adams}.
Here in Figure \ref{fig:E_2SU5}, $p=2$, we can read from the Adams chart
$\Omega^{\Spin \times \SU(5)}_0=\Z$ (an infinite line),
$\Omega^{\Spin \times \SU(5)}_1=\Z_2$ (a dot),
$\Omega^{\Spin \times \SU(5)}_2=\Z_2$ (a dot),
$\Omega^{\Spin \times \SU(5)}_3=0$ (nothing),
$\Omega^{\Spin \times \SU(5)}_4=\Z^2$ (two infinite lines),
$\Omega^{\Spin \times \SU(5)}_5=0$ (nothing),
and
$\Omega^{\Spin \times \SU(5)}_6=0$ (an infinite line).

\subsubsection{Classification of all invertible anomalies of ${\Spin \times \SU(5)}$ fermion theories}

By \Eq{eq:Adams} and \Eq{eq:Pontryagin-Thom}, we obtain the bordism group $\Omega^{\Spin \times \SU(5)}_D$ shown in Table \ref{table:SU5Bordism}, focusing on $D=4,5,6$.
\begin{table}[!h]
\centering
\begin{tabular}{ c c c}
\hline
\multicolumn{3}{c}{Bordism group}\\
\hline
$D$ & 
$\Omega^{\Spin \times \SU(5)}_D$
& generators \\
\hline
4 & $\Z^2$ & $\frac{\sigma}{16},c_2$\\
5 & $0$ \\
6 & $\Z$ & $\frac{c_3}{2}$ \\
\hline
\end{tabular}
\caption{Bordism group $\Omega^{\Spin \times \SU(5)}_D$. 
$\sigma$ is the signature of manifold.
{The $c_j$ is the $j$-th Chern class of the associated vector bundle of $\SU(n)$.}
Note that $c_3=\Sq^2c_2=(w_2+w_1^2)c_2=0\mod2$ on Spin 6-manifolds.
Actually $\Omega^{\Spin \times \SU(n)}_D=\Omega^{\Spin \times \SU(n+1)}_D$ for $n\ge3$ and $0\le D\le 6$.
}
\label{table:SU5Bordism}
\end{table}

By \Eq{eq:Ext-extension} and \Eq{eq:L-extension}, 
we obtain the cobordism group $\TP_D(\Spin \times \SU(5))$ shown in Table \ref{table:SU5TP}, focusing on $D=4,5$.

\begin{table}[!h]
\centering
\begin{tabular}{ c c c}
\hline
\multicolumn{3}{c}{Cobordism group}\\
\hline
$d$ & 
$\TP_D(\Spin \times \SU(5))$
& generators \\
\hline
4 & $0$ \\
5 & $\Z$ & $\frac{1}{2}$CS$_5^{\SU(5)}$  \\
\hline
\end{tabular}
\caption{Topological phase classification ($\equiv$ TP) as a cobordism group $\TP_D(\Spin \times \SU(5))$, following Table \ref{table:SU5Bordism}. 
}
\label{table:SU5TP}
\end{table}


\subsection{Cobordism groups and topological phases for $\frac{\Spin \times
\Spin(10)}{{\Z_2^f}}$
and  $\frac{\Spin \times
\Spin(18)}{{\Z_2^f}}$:\\ 
SU(10) and SU(18) Grand Unification}


We consider $G=\frac{\Spin \times \Spin(10)}{\Z_2^F}$
for the  Fritzsch-Minkowski SO(10) Grand Unification \cite{FM7593}, 
the Thom-Madsen-Tillmann spectrum $MTG$  of the group $G$ is
\bea
MTG=M\Spin\wedge\Sigma^{-10}M\SO(10).
\eea
The $T$ in $MTG$ means the $G$-structures are on tangent bundles instead of normal bundles.
In this case, we have 
$w_2(TM)=w_2(V_{\SO(10)})$.

For the dimension $D=t-s<8$, since there is no odd torsion (see the footnote \ref{ft:no-odd-torsion}), by 
$MTG=M\Spin\wedge X$, then 
$\pi_D(MTG)=\pi_D(ko\wedge X)$ for $D<8$;
so for the dimension $D=t-s<8$, 
from \Eq{eq:Adams}, we have
\bea
\Ext_{\A_2(1)}^{s,t}(\H^*(X,\Z_2),\Z_2)\Rightarrow(\Omega_{D=t-s}^G)_2^{\wedge}.
\eea
Hence, we have the Adams spectral sequence
\bea
\Ext_{\A_2(1)}^{s,t}(\H^{*+10}(M\SO(10),\Z_2),\Z_2)\Rightarrow\Omega_{D=t-s}^{\frac{\Spin \times \Spin(10)}{\Z_2^F}}. \quad\quad\quad 
\eea

\begin{figure}[!h]
\begin{center}
\includegraphics[scale=.7]{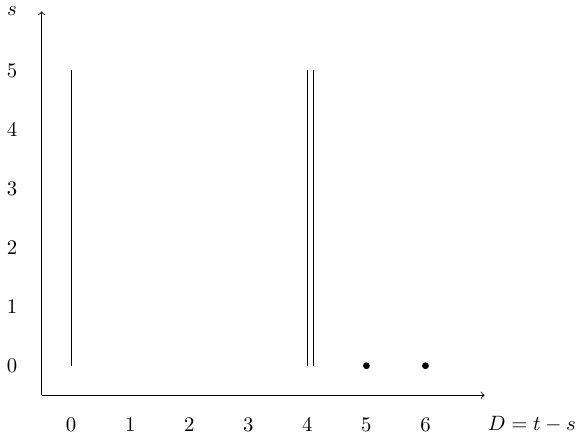}
\end{center}
\caption{Adams chart for  $\Omega_D^{\frac{\Spin \times \Spin(10)}{\Z_2^F}}$, also for 
$\Omega_D^{\frac{\Spin \times \Spin(18)}{\Z_2^F}}$.}
\label{fig:E_2MSO10}
\end{figure}

Actually we find \cite{WW1910.14668}
$$
\Omega^{\frac{\Spin \times \Spin(10)}{\Z_2^F}}_D=
\Omega^{\frac{\Spin \times \Spin(18)}{\Z_2^F}}_D=
\Omega^{\frac{\Spin \times \Spin(n)}{\Z_2^F}}_D=\Omega^{\frac{\Spin \times \Spin(n+1)}{\Z_2^F}}_D,
$$ 
for $n\ge7$ and $0\le D\le 6$.

The $\A_2(1)$-module structure of $\H^{*+10}(M\SO(10),\Z_2)$ below degree 6 is shown in \Ref{WW1910.14668}'s Sec.~6's Figure 27,
and the $E_2$ page is shown in Figure \ref{fig:E_2MSO10}.
Here we have used the correspondence between $\A_2(1)$-module structure and the $E_2$ page shown in 
Appendix A
of \Ref{WW1910.14668}.

To read the result from the Adams chart Figure \ref{fig:E_2MSO10}, we look at the stabilized
$E_{\infty}$ page, one dot indicates a finite group $\Z_p$, 
a vertical finite line segment connecting $n$ dots indicates a finite group $\Z_{p^n}$. 
But when $n=\infty$, the infinite line connecting infinite dots indicates an infinite group, an integer $\Z$. 
Here $p$ is given by the mod $p$ Steenrod algebra $\A_p$ in \Eq{eq:Adams}.
Here in Figure \ref{fig:E_2MSO10}, $p=2$, we can read from the Adams chart
$\Omega^{\frac{\Spin \times \Spin(10)}{\Z_2^f}}_0=\Z$ (an infinite line),
$\Omega^{\frac{\Spin \times \Spin(10)}{\Z_2^f}}_1=0$ (nothing),
$\Omega^{\frac{\Spin \times \Spin(10)}{\Z_2^f}}_2=0$ (nothing),
$\Omega^{\frac{\Spin \times \Spin(10)}{\Z_2^f}}_3=0$ (nothing),
$\Omega^{\frac{\Spin \times \Spin(10)}{\Z_2^f}}_4=\Z^2$ (two infinite lines),
$\Omega^{\frac{\Spin \times \Spin(10)}{\Z_2^f}}_5=\Z_2$ (a dot),
and
$\Omega^{\frac{\Spin \times \Spin(10)}{\Z_2^f}}_6=\Z_2$ (a dot).

\subsubsection{Classification of all invertible anomalies of 
$\frac{\Spin \times
\Spin(10)}{{\Z_2^f}}$ and  
$\frac{\Spin \times
\Spin(18)}{{\Z_2^f}}$
fermion theories}

By \Eq{eq:Adams} and \Eq{eq:Pontryagin-Thom}, we obtain the bordism group 
$\Omega^{\frac{\Spin \times \Spin(10)}{\Z_2^f}}_D$ shown in Table \ref{table:Spin10Bordism}, 
focusing on $D=5$.


\begin{table}[!h]
\centering
\begin{tabular}{ c c c}
\hline
\multicolumn{3}{c}{Bordism group}\\
\hline
$D$ & 
$\Omega^{\frac{\Spin \times \Spin(10)}{\Z_2^f}}_D$
& generators \\
\hline
5 & $\Z_2$ & {$\w_2(TM) \w_3(TM)=\w_2(V_{\SO(10)}) \w_3(V_{\SO(10)})$} \\
\hline
\end{tabular}
\caption{Bordism group. 
The same result holds for $\Omega^{\frac{\Spin \times \Spin(18)}{\Z_2^F}}_D$ and
$\Omega^{\frac{\Spin \times \Spin(n)}{\Z_2^F}}_D$ with $n\ge7$ and $0\le D\le 6$.
}
\label{table:Spin10Bordism}
\end{table}

By \Eq{eq:Ext-extension} and \Eq{eq:L-extension}, 
we obtain the cobordism group $\TP_D({\frac{\Spin \times \Spin(10)}{\Z_2^f}})$ shown in Table \ref{table:Spin10TP}, focusing on $D=5$.

\begin{table}[!h]
\centering
\begin{tabular}{ c c c}
\hline
\multicolumn{3}{c}{Cobordism group}\\
\hline
$D$ & 
$\TP_D({\frac{\Spin \times \Spin(10)}{\Z_2^f}})$
& generators \\
\hline
5 & $\Z_2$ & {$\w_2(TM) \w_3(TM)=\w_2(V_{\SO(10)}) \w_3(V_{\SO(10)})$} \\
%
\hline
\end{tabular}
\caption{Topological phase classification ($\equiv$ TP) as a cobordism group, following Table \ref{table:Spin10Bordism}. 
Same result for $\TP_D({\frac{\Spin \times \Spin(18)}{\Z_2^F}})$ and
$\TP_D({\frac{\Spin \times \Spin(n)}{\Z_2^F}})$ with $n\ge7$  and $0\le D\le 5$.
}
\label{table:Spin10TP}
\end{table}



\bibliography{./all,./publst,./local,./JWnewSO10}

\end{document}